\documentclass[twocolumn]{aastex631}
\usepackage{amsmath}

\begin{document}

\title{Mainly on the Plane: Observing the Extended, Ionized Disks of Milky Way Analogs in IllustrisTNG}
\shorttitle{Observing the Extended, Ionized Disks of MW Analogs in IllustrisTNG}

\author[0009-0003-3855-2708]{Michael Messere}
\affiliation{Department of Astronomy, Columbia University, New York, NY 10027, USA}
\author[0000-0003-0789-9939]{Kirill Tchernyshyov}
\affiliation{Department of Astronomy, University of Washington, 
Seattle, WA 98195, USA}

\author[0000-0002-1129-1873]{Mary E. Putman}
\affiliation{Department of Astronomy, Columbia University, New York, NY 10027, USA}
\author[0000-0003-2630-9228]{Greg L. Bryan}
\affiliation{Department of Astronomy, Columbia University, New York, NY 10027, USA}
\author[0000-0002-0355-0134]{Jessica K. Werk}
\affiliation{Department of Astronomy, University of Washington, 
Seattle, WA 98195, USA}

\author[0000-0003-4158-5116]{Yong Zheng}
\affiliation{Department of Physics, Applied Physics and Astronomy, Rensselaer Polytechnic Institute, Troy, NY 12180, USA}

\author[0000-0003-2666-4430]{David Schiminovich}
\affiliation{Department of Astronomy, Columbia University, New York, NY 10027, USA}

\correspondingauthor{Michael Messere}
\email{mam2645@columbia.edu}

\begin{abstract}

This paper explores the extent to which the circumgalactic medium (CGM) of Milky Way-like galaxies is located in an extended, ionized, disklike structure. To test this hypothesis, we analyze the spatial and kinematic distributions of different ion species within a sample of MW-like systems in IllustrisTNG. We model commonly observed ions (H $\textsc{i}$, Mg $\textsc{ii}$, Si $\textsc{iv}$, C $\textsc{iv}$, and O~$\textsc{vi}$) and calculate (1) their angular momentum misalignment from the star-forming disk ($\theta$) and (2) the fraction of absorption consistent with galaxy rotation ($f_\mathrm{EWcorot}$). We find that 63\% of Mg~\textsc{ii}, 45\% of Si~\textsc{iv}, 38\% of C~\textsc{iv}, and 35\% of O~\textsc{vi} mass along the major axis have kinematics aligned with the galaxy angular momentum axis.  We extend this to a mock absorption line survey and quantify $f_\mathrm{EWcorot}$. We find that $f_\mathrm{EWcorot}$(Mg~\textsc{ii}) $\sim80\%$ and  $f_\mathrm{EWcorot}$(O~\textsc{vi}) $\sim60\%$ at $\sim0.5\ \mathrm{R_{200c}}$, in agreement with recent observational work. We find that in the typical MW analog, there is evidence of cool-warm material in an extended, corotating structure, regardless of whether the angular momentum or observational definition is used.  Hence, we expect that the typical MW CGM, especially in the low ions, \emph{should} be mainly on the plane.

\end{abstract}

\section{Introduction} \label{sec:intro}

The idea of a complex, multi-phase medium surrounding the interstellar medium (ISM), often referred to as the circumgalactic medium (CGM), can be traced to the midtwentieth century when a diffuse, hot corona surrounding our Milky Way (MW) was first proposed \citep{Spitzer1956}. As Lyman Spitzer pointed out, it seems as though an ionized medium composed of cool, warm, and hot components $\emph{should}$ be everywhere we point our instruments.  It was subsequently proposed that nearly all absorption lines in external, normal galaxies are the result of a similar extended, gaseous halo \citep{Bahcall1969}. However, it wasn’t until the launch of the $\emph{Hubble Space Telescope}$ (HST)  \citep{Lanzetta1995,Chen1998} and its \emph{Cosmic 
Origins Spectrograph} (COS) \citep{Green2012,Tumlinson2011,Tumlinson2013,Werk2013,Bordoli2014,Werk2014,Werk2016,Bordoloi2018} that the CGM could be explored in detail from an observational point-of-view, especially the spatial and kinematic extent of the gas. 

The radial extent of the gaseous disk has been a long-standing question, especially as it relates to how gas accretes onto a galaxy \citep{Sancisi2008,Fox_Dave2017}. Over the past decade, there has been mounting evidence for an extended rotating disk in external, edge-on L$^\ast$ galaxies. This has been explored in emission, connecting H~\textsc{i}, H$\alpha$, and the metagalactic UV background \citep{Maloney1993,Bland-Hawthorn1997,Adams2011,Fumagalli2017,Bland-Hawthorn2017}. In addition, there has been extensive absorption work, where individual sightlines (e.g., Mg~\textsc{ii}, O~\textsc{vi}) have been used to probe the spatial distribution and kinematics of the ionized CGM  \citep{Bordoloi2011,Kacprzak2012,Bordoloi2014_2,Kacprzak2015,Ho2017,Martin2019,French_Wakker2020,nateghi24,Nateghi2024_2,Ho2025,Kacprzak2025}. For example, a sample of galaxy–quasar pairs from \cite{Ho2017} showed evidence of corotating Mg~\textsc{ii} out to nearly 100 kpc ($\sim 0.5\ \mathrm{R_{vir}}$) along the midplane. More recently, \cite{Ho2025} found that O~\textsc{vi} is more likely to co-rotate with the disk when there is a low-ion match (e.g., Mg~\textsc{ii}), suggesting that the low and high ions are co-spatial. \cite{Nateghi2024_2} (see also \cite{nateghi24,Kacprzak2025}) calculated the absorption equivalent width co-rotation fraction ($f_\mathrm{EWcorot}$), defined as the fraction of absorption consistent with galaxy rotation. They found evidence for stronger co-rotation in lower ionization gas and found that highly corotating O~\textsc{vi} is found preferentially along the major axis. 

Together, this work has demonstrated that there is a substantial amount of ionized gas out to large radii with evidence of co-rotation. However, the single sightline approach of absorption line spectroscopy has made it difficult to develop a complete picture of this extended, corotating disk and its connection to the inner, star-forming disk is still unclear. In order to aid in the interpretation of these findings, astronomers have turned to the recent advancement in computation, which has enabled the detailed study of the CGM at various scales: from the interaction between the cold and hot phase at the parsec level \citep{McCourt2018,Gronke2020,Fielding2020} to large-scale cosmological simulations that explore different star-formation and feedback models \citep{Hopkins2014,Schaye2015,Crain2015,Tremmel2017,Nelson2019,Peeples2019,Dave2019,Smith2024,Villaescusa-Navarro2021}. 

From the simulation perspective, there has been considerable interest in modeling the CGM and the connection between gas accretion, feedback, and disk-formation \citep[e.g., ][]{Hummels2013,Ford2013,Ford2014,Ford2016,Liang2016,Nelson2018,Zheng2019,Hummels2019,Peeples2019,DeFelippis2020,Ho2020}. Cosmological simulations have shown that cold-mode gas accretion aids in disk-formation and is often corotating with the disk \citep{Dekel2009,Stewart2011,Stewart2011b}. Efforts to bridge simulation and observation have included studying the angular momentum of cold gas \citep{DeFelippis2020} and mock absorption-line studies \citep{Ho2019,DeFelippis2021}, both of which suggest that corotating cold gas along the midplane can extend well beyond the optical and emission line-detected disk -- consistent with observations.

Despite this extensive work looking outside the MW, recent observational work has ironically suggested that we don't fully understand our own backyard -- the same backyard that Lyman Spitzer pointed out. In particular, there is recent observational evidence that our CGM is a possible anomaly compared to the massive and extensive CGM of L$^*$ galaxies \citep{Prochaska2011,Zheng2019,Wilde2021,Wilde2023}. Through the use of both distant quasars and bright halo stars at high Galactic latitude, it has been shown there is a substantial amount of ionized material within 10 kpc of the mid-plane \citep{Sembach2003,Savage2003,Savage_Wakker2009,Lehner_Howk2011,Werk2019}. There has also been recent discussion about the extent of this gas beyond 10 kpc, especially since the low-velocity halo gas blends with the inner disk gas \citep{Zheng2015}. The QuaStar Survey \citep{Bish2021} demonstrated that a majority of the ionized gas (as traced using C~\textsc{iv}) is in fact confined within a vertical distance of $\sim$ 10 kpc from the Galactic midplane. These results lend credence to the hypothesis that the Milky Way harbors anomalously low levels of ionized gas compared to observations of external galaxies (see their Figure 6). Although it is possible that the MW-CGM is different, it is important to consider the alternative possibility that a significant component of the ionized gas resides in an extended, rotating disk.  This would be consistent with the co-rotation already identified in both observational and simulation work.  

This is also suggested in \cite{Qu2022}, where 265 quasar and stellar sightlines were used to probe C~\textsc{iv} in the MW. They found two possible solutions in their analysis, including an extended disk in the radial direction. However, since gas in an extended disk is largely obscured by the MW ISM, more data is needed. In addition, the velocity and distance of the gas in absorption is difficult to model, therefore requiring simulations to help aid in their interpretation.  In simulation work, there has been little effort connecting co-rotation to ions that trace the present-day MW, including the Si~\textsc{iv} and C~\textsc{iv} detected using HST/COS.     

In this work, we aim to address the latter, and use a sample of 88 MW analogs from the IllustrisTNG simulation to study the spatial and kinematic distribution of halo gas, traced using H $\textsc{i}$, Mg $\textsc{ii}$, Si $\textsc{iv}$, C $\textsc{iv}$, and O $\textsc{vi}$. We quantify the gas alignment relative to the angular momentum axis of the star-forming disk ($\theta$), an attempt to bridge both observational and simulation work. In this paper, we are interested in the following:

\begin{enumerate}
    \item What is the angular momenta distribution the cool and warm halo gas (traced using H $\textsc{i}$, Mg $\textsc{ii}$, Si $\textsc{iv}$, C $\textsc{iv}$, and O $\textsc{vi}$) relative to the angular momentum axis of the star-forming disk? 
    
    \item For a given ion, how is the angular-momentum-aligned column density and mass distributed throughout the galaxy relative to the total column density and mass? 

    \item How does our definition of the angular momentum misalignment angle ($\theta$) translate into a mock absorption line study, particularly when gas observability is taken into account?

    \item What is the \emph{corotating} fraction ($f_{\mathrm{EWcorot}}$) of our sample, as defined in recent observational work? 
\end{enumerate}  

Our analysis offers a complementary, external perspective to the ongoing Mainly on the Plane collaboration \citep{Werk2021hst,Werk2024}, which explores the same question from an observational point-of-view from within the Galaxy with a sample of low-latitude quasars. Together, these efforts aim to determine if the MW has an extended corotating disk.

The paper is organized as follows. We first present our sample of MW analogs in IllustrisTNG in \S\ref{sec:defining the sample}. We then define angular momentum alignment in \S\ref{sec:methods} and compare an idealized disk galaxy with our cosmological sample. We next present our results in \S\ref{sec:results}, where we quantify the spatial extent of the angular momentum misalignment angle ($\theta$) and co-rotation fraction ($f_\mathrm{EWcorot}$). We then connect our work to observational and computational work in \S\ref{sec:discussion}. We also discuss the role of satellites in driving the angular momentum alignment and the caveats of using IllustrisTNG in \S\ref{sec:discussion}. Finally, we summarize our findings and conclude in \S\ref{sec:conclusion}. 

\section{TNG MW-like Galaxy Selection}
\label{sec:defining the sample}

In this paper, we focus on a selection of MW-like galaxies taken from TNG50-1 \citep{Nelson2019,Nelson2019a}, the highest resolution installment of the publicly available IllustrisTNG project ($2 \times 2160^3$ resolution elements across the $\sim 50$ comoving Mpc per side volume). The different components (e.g., dark matter, gas, stars) of the simulation are evolved from $z = 127$ to $z = 0$ using \textsc{arepo}, a cosmological magnetohydrodynamical moving-mesh code \citep{Springel2010,Ruediger2011}. The simulation adopts an average baryon cell mass of $8.5 \times 10^4\, \mathrm{M}_{\odot}$ and an average cell size of $70 - 140$ pc in the star-forming regions. This resolution is comparable to many zoom-in simulations, but with the additional benefit of a large sample size – hence our initial motivation in using TNG50 data for this study. 

\begin{figure}
\includegraphics[width=\linewidth]{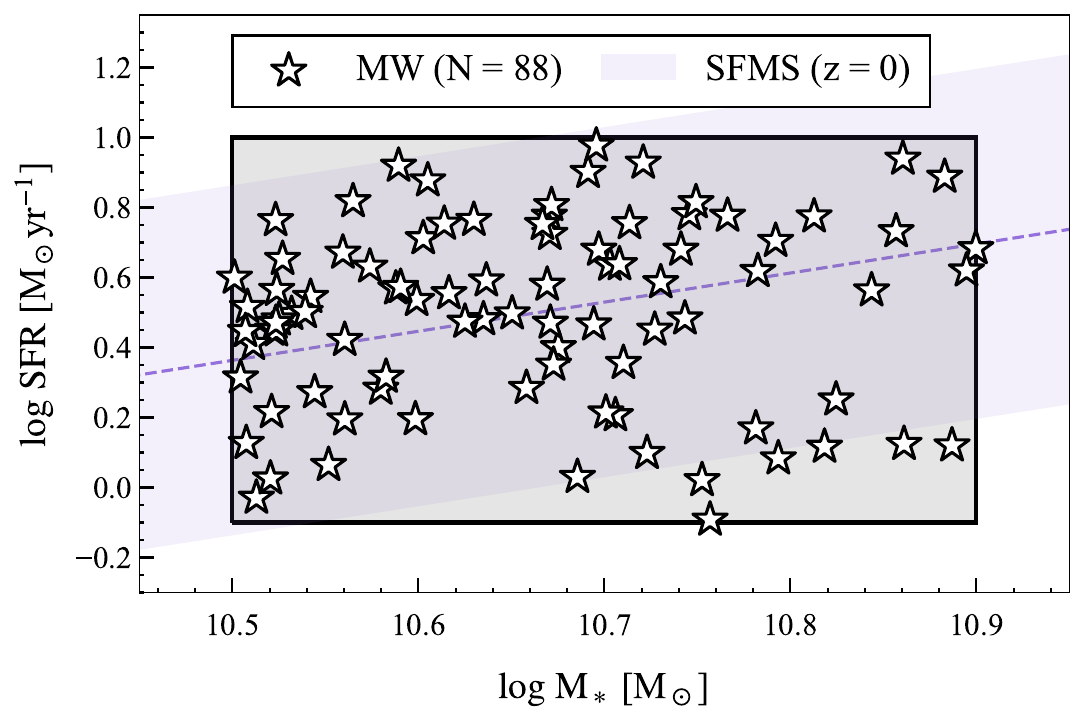}
\caption{The $\mathrm{M_\ast \mbox{-} SFR}$ selection of TNG-50 MW analogs adopted from \cite{Pillepich2023}. The MW selection region (black box) is bounded by star-formation rate and mass estimates of the MW, as defined in \cite{Pillepich2023}. The present-day TNG50 star-forming main-sequence (purple) is included for reference. 
\label{fig:selection}}
\end{figure}

Since we are interested in the distribution of gas around disklike galaxies similar to our own MW, we start with the same selection criteria described in \citet{Pillepich2023}. The sample can be found on their public website\footnote{\url{https://www.tng-project.org/data/milkyway+andromeda/}} and systems within this sample have been used in other work exploring MW analogs in TNG50, including MW disk-formation \citep{Pillepich2019,Semenov2024b,Semenov2024c}, gas morphology and angular momentum \citep{Kauffmann2019,DeFelippis2020,Waters2024}, and the CGM \citep{Ramesh2023,Ramesh2023b,Ramesh2024}. The authors define a sample of 198 MW/M31-like galaxies based on stellar mass, morphology, and environment at z = 0. The requirements are: \\

\begin{enumerate}
    \item \textbf{Stellar Mass:}
        The stellar mass within 30 kpc is within the range $\mathrm{\log M_{*}/M_\odot(<30\:kpc) = 10.5-11.2}$. 
    \item \textbf{disklike:}
        The minor-to-major axis ratio of the stellar mass distribution is less than 0.45 ($\mathrm{c \big/ a\leq 0.45}$) \emph{or} the system appears disklike with spiral arms through visual inspection.
    \item \textbf{Isolated System:}
        There is no galaxy with $\mathrm{\log M_{*}/M_\odot \geq 10.5}$ within 500 kpc (M31 is $\sim$ 785 kpc) \emph{and} $\mathrm{M_{200c}(host) < 10^{13}\:M_\odot}$.
\end{enumerate}  

See Section 2.3 of \citet{Pillepich2023} for justification of these MW/M31 selection criteria. We further refine this sample of 198 MW/M31 systems. First, we use the same MW $\mathrm{M_\ast-SFR}$ selection presented in \citet{Pillepich2023} (see their Figure 13) since we are interested in MW analogs and not the more massive, less star-forming M31 analogs. We approximate these as $\mathrm{\log M_\ast/M_\odot = 10.5 - 10.9}$ (within 30 kpc) and $\log \mathrm{SFR/(M_\odot\ yr^{-1})} = -0.1 - 1.0$ (50 Myr. average). In addition, we ignore any system that appears as an S0 galaxy (disky stellar morphology but no spiral arms present), is a non-central subhalo, or is located near the edge of the cosmological box (only a few systems). 

Our final sample of 88 is shown in Figure \ref{fig:selection}. For reference, the star-forming main-sequence (SFMS) for TNG50 is included as the purple line (and $\pm0.5\, \mathrm{dex}$). 
It is also important to note that satellites are not included in our analysis because we are only interested in the angular momentum offset angle distribution of gas directly associated with the MW analog. Specifically, we use the subhalo cutouts, where gravitationally bound substructures are assigned to the same subhalo (see Section 3 in \cite{Nelson2019a}). We explore the role of satellites in \S\ref{subsec:satellites}. 

\section{Methodology}
\label{sec:methods}

With our MW analog sample defined, we now present the methods used to model the ions and calculate the angular momentum and co-rotation of the CGM gas. We first discuss modeling the ions in \S\ref{subsec:trident} and then define the angular momentum misalignment angle in \S\ref{subsec:define misalignment angle} and \S\ref{subsec:agora}. We lastly discuss how we calculate observables, including column density maps in \S\ref{subsec:column_density_corotation_fraction} and mock absorption sightlines in \S\ref{subsec:sightline_generation}.
\subsection{Calculating Ion Species using \texttt{TRIDENT}}\label{subsec:trident}

In order to approximate the number density of ion species not available in the output of TNG50 (Mg~$\textsc{ii}$, Si~$\textsc{iv}$, C~$\textsc{iv}$, and O~$\textsc{vi}$), we use the open-source, simulation post-processing tool \texttt{TRIDENT} \citep{trident}. The H~$\textsc{i}$ abundance for each gas cell, and individual abundances of H, He, C, N, O, Ne, Mg, Si, and Fe are followed in IllustrisTNG. \texttt{TRIDENT} provides a look-up table of ion abundances for a given density and temperature assuming collisional ionization equilibrium and photoionization from a metagalactic ultraviolet background. \texttt{CLOUDY} \citep{cloudy} and \texttt{ROCO} \citep{ROCO} are used to calculate this table using nearly the same hydrogen density and temperature range found in TNG50. We use the same ultraviolet background (with self-shielding) that TNG50 was evolved with \citep{FG2009}. We note that we do not include the contribution of the radiation field from the central galaxy itself. We show (Appendix~\ref{appendix:stellar radiation field}) that the increase in the radiation field doesn't significantly change our results since the slope of a galactic radiation field is similar to the ultraviolet background, especially within the ionization energy range of Mg~$\textsc{ii}$, Si~$\textsc{iv}$, C~$\textsc{iv}$, and O~$\textsc{vi}$ (see Section 6.3 in \cite{Liang2016} for a similar discussion). In addition, we refrain from using star-forming gas in the remainder of this analysis since there is a non-physical temperature used in the effective equation of state two-phase ISM sub-grid model \citep{Springel2003}.
 
For a given non-star-forming gas cell hydrogen density and temperature, \texttt{TRIDENT} calculates the following:
\begin{equation}\label{Equation1}
    n_{X_i} = n_X f_{X_i}
\end{equation} 
where $n_X$ is the total number density of an element X and $f_{X_i}$ is the ionization fraction of the $i$'th ion of the element. This is repeated for each gas cell in the halo. 

\subsection{Defining the Misalignment Angle}
\label{subsec:define misalignment angle}

For each MW analog cutout within our TNG50 sample, we are interested in the angular momentum misalignment of the CGM gas relative to the star-forming disk. There are several approaches in the literature for defining co-rotation. In absorption-line observational work, a sightline is often considered corotating when the sign of the ion equivalent-width-weighted mean line-of-sight velocity matches the sign of the inner disk projected rotation velocity \citep[e.g.,][]{Martin2019}. In comparison, simulations have three-dimensional kinematic information available for each gas cell. In a sample from the \texttt{FIRE} project, \citet{El-Badry2018} calculated the orbital circularity parameter ($\epsilon$) of each gas particle (relative to the specific angular momentum of the H~$\textsc{i}$ in the galaxy) and found that a majority of gas resides at $\epsilon \geq 0.70$. This is similarly computed for the stellar component of TNG50 MW analogs in \citet{Emami2021}, where they defined the stellar disk as being composed of star particles where $\epsilon \geq 0.70$. \cite{Sales2012} quantify $\epsilon$ to describe disk- versus bulge-dominated systems, but also introduced the misalignment angle $\theta$ between the angular momentum vector within a given enclosed mass fraction and the total spin of the system. In this work, we adopt a similar definition of corotating gas to \cite{Sales2012}, but we calculate the angular momentum misalignment angle, $\theta$, for each gas cell.

First, we calculate the angular momentum axis of the star-forming gas disk of each galaxy, defined as gas cells within 30 kpc of the galactic center. We do not select a gas density threshold, although the calculation is more heavily weighted toward star-forming gas ($n_\mathrm{H} \gtrsim 0.1\ \mathrm{cm}^{-3}$). In TNG MW analogs with high angular momentum, it has been shown that the CGM angular momentum vector can be misaligned from the stellar disk angular momentum vector by $\sim 15^\circ$  \citep{DeFelippis2020}. This is a relatively small value and therefore we don't expect a large discrepancy relative to observational work that uses stellar photometry to define inclination \citep[e.g.,][]{Ho2017}. We define the center of each cutout as the gas cell with the lowest potential energy and subtract off the bulk motion of the central galaxy from each gas cell. The specific angular momentum axis of the gas disk is therefore:     
\begin{equation}\label{Equation:jdisk}
    \vec{j}_{disk} = \frac{1}{M_{disk}} \sum_{i = 1}^{N} m_i \big(\vec{r}_i - \vec{r}_{center}\big) \times \big(\vec{v}_i - \vec{v}_{bulk}\big)
\end{equation} 
where $\mathrm{M_{disk} = \sum_{i = 1}^{N} m_i}$ is the total gas mass of the disk region, $\vec{r}_{\mathrm{center}}$ is the position of the most bound particle and $\vec{v}_{\mathrm{bulk}}$ is the bulk velocity of the central galaxy. The bulk velocity is calculated using both the gas and particle data within the entire subhalo.  

Since we are interested in the kinematic behavior of the ionized gas in the CGM relative to the star-forming gas disk, we define the angular momentum misalignment angle ($\theta$) between each gas cell and the star-forming gas disk. The specific angular momentum axis for each gas cell is calculated as: 
\begin{equation}\label{Equation1.5}
    \vec{j}_{i} = \vec{r}_{i\mathrm{,rot.,z=0}} \times \vec{v}_{i\mathrm{,rot.}}
\end{equation}
where $\vec{r}_{i\mathrm{,rot.,z=0}}$ and $\vec{v}_{i\mathrm{,rot.}}$ represent the position and velocity in the frame rotated such that the z-axis is aligned with the disk angular momentum axis. We also set z = 0 for all gas cells, since we will not find a misalignment angle less than the gas cell galactic latitude angle \emph{if} we use the three-dimensional position. The angular momentum misalignment angle is given by: 

\begin{equation}\label{Equation2}
    \theta = \arccos{\big(\hat{j}_{i} \cdot \hat{z})}\\
\end{equation}
where $\hat{j}_{i} = \frac{\vec{j}_{\mathrm{i}}}{|\vec{j}_{i}|}$. 

It is possible that some of the gas is aligned with the angular momentum axis by chance alone. For example, \cite{Ho2017} discuss how a spherical distribution of random cloud orbits will result in a galaxy-aligned rotational velocity half of the time (see their Section 3.1). In order to quantify the role of random motion, we produce a random distribution of angular momentum vectors by uniformly sampling points on the surface of a sphere twice, representing random position and velocity vectors. We use $\phi=\mathrm{2\pi\ \cdot N}$ and $\theta=\mathrm{\cos^{-1}\big(1-2\ \cdot N
\big)}$, where N is $10^6$ random samples drawn from a uniform distribution over [0, 1). We then convert these spherical coordinates to Cartesian coordinates and calculate the component of their cross product aligned with the z-axis, similar to Equation \ref{Equation2}. This sample is hereafter labeled as  \lq random motion\rq{} in our analysis. We also compared our definition of the misalignment angle with the orbital circularity parameter ($\epsilon \geq 0.70$) and found that they select nearly the same disklike structure.
\subsubsection{Test Case: The AGORA Galaxy}\label{subsec:agora}

In order to test our definition of the angular momentum misalignment angle, we use an idealized simulation of a MW-like galaxy from the Assembling
Galaxies Of Resolved Anatomy (AGORA) project \citep{Kim_2016,Goldbaum2015,Goldbaum2016}. The fiducial simulation at $\mathrm{z\sim 0}$ with feedback we used is publicly available\footnote{\url{https://girder.hub.yt/\#collection/573647d3dd9119000164acf0}}. Since there is no cosmological gas accretion, we naively expect almost all of the gas to be corotating. 
In fact, we find that the bulk ($\sim 97 \%$) of the gas cells have a misalignment angle $\theta \leq 30^\circ$. The remaining $\sim 3 \%$ of the gas has some other motion relative to the galaxy angular momentum axis. For example, there is feedback-driven outflowing material, likely having $\theta \sim 90^\circ$. Our investigation into the idealized MW disk simulation demonstrates that $\theta \leq 30^\circ$ is an effective definition of gas corotating with a galaxy's disk.

Hereafter, we apply $\theta \leq 30^\circ$ to our IllustrisTNG MW analogs to investigate the degree of misalignment between CGM and disk gas in cosmological simulations. In this paper, we use angular momentum \emph{misalignment} to refer to the distribution of gas and angular momentum \emph{alignment} to refer to the subset of gas aligned with the star-forming disk angular momentum axis ($\theta \leq 30^\circ$). In addition, we reserve the term corotating for the line-of-sight velocity definition often used in observational work. 

\subsection{Column Density \& Angular Momentum Alignment Map}\label{subsec:column_density_corotation_fraction}

Throughout this work, we present the total column density for each ion, $\mathrm{N(\theta\leq180^\circ)}$. In addition, we present the sub-component of the total column density that is aligned, $\mathrm{N(\theta\leq30^\circ)}$. To calculate both the total and aligned column density,  we use \textsc{yt} to create a projection. For the aligned column density, we select halo gas where $\theta\leq30^\circ$. We normalize the physical distance in each projection by $\mathrm{R_{200c}}$ such that each projection is $\mathrm{2\ R_{200c}}$ across. The projection is centered on the gas cell with the lowest gravitational potential energy. The resolution of a pixel in each projection is $\mathrm{(0.002\ R_{200c})^2}$. For $\mathrm{\bar{R}_{200c}\sim215\ kpc}$, this is equivalent to $\mathrm{(0.5\ kpc)^2}$. The TNG50-1 resolution in the CGM is coarser than this resolution \citep[$\sim2$ kpc at $0.50\ \mathrm{R_{200c}}$, see][]{Ramesh_Nelson2024}; however, we do not find a significant change in our result if we adjust the resolution. 

In the edge-on projection, we orient the projection such that the angular momentum axis of the inner galaxy is along the y-axis and the disk plane is along the x-axis.  Hereafter, we will refer to the y-axis as the minor axis and x-axis as the major axis. We rotate the edge-on projection in increments of $15^\circ$ until it is face-on. We repeat this for each ion and the total and aligned column density. In total, there are 70 projections for each galaxy in our sample.     

We then decompose each projection to construct column density profiles as a function of (i) radial distance for inclination-dependent profiles, (ii) distance along the major axis for major axis profiles, or (iii) distance along the minor axis for minor axis profiles. For profiles as a function of inclination, we use the entire projection. For major axis profiles, we use the edge-on projection and select gas within $\pm0.1\,\mathrm{R_{200c}}$ of the disk plane. For minor axis profiles, we similarly use the edge-on projection and select gas within a cylinder aligned with the minor axis and extending to a radius of $\mathrm{R_{200c}}$. For each galaxy, we use the median column density, since this is representative of the typical column density a sight-line will pass through. This procedure is repeated for each ion, and for both the total column density and its aligned sub-component.

We also compute the ratio of the aligned column density to the total column density ($\mathrm{N(\theta\leq30^\circ)}\,/\,\mathrm{N(\theta\leq180^\circ)}$) in each projection, yielding a two-dimensional map of the aligned fraction. We then take the median across these individual maps. We note that since we only use one edge-on image for each galaxy, there is some level of patchiness, even after computing the median across the sample. Ideally, each galaxy is randomly re-oriented at the same inclination in order to smooth out some of these features. Instead, a Gaussian filter is applied to the contours using the image processing tool \texttt{scipy.ndimage} with a bandwidth of $\sigma=20$ pixels ($\mathrm{0.04\ R_{200c}}$). This choice of $\sigma$ shows the general spatial structure of each ion, without including too much noise (small $\sigma$) or washing out any disklike and filamentary features (large $\sigma$). 

\subsection{Sightline Generation}\label{subsec:sightline_generation}

The definition of angular momentum misalignment adopted in this paper is difficult to connect to what is observed in the real Universe (e.g., using HST/COS). These surveys often rely on pencil-beam quasar sight-lines that are unable to probe the same extent as a cosmological simulation. How does our definition of angular momentum alignment relate to what would be observed? In order to produce mock HST/COS absorption line observations, we use \textsc{trident}. In addition to providing a look-up table to estimate the ionization fraction, \textsc{trident} can estimate the absorption signature along a sight-line using the line-of-sight velocity and ionization fraction for a given ion along a pre-defined ray. 

We produce a grid of sight-lines that is aligned with the edge-on view of each galaxy. The grid spans $\mathrm{\pm 0.75\ R_{200c}}$ along the major axis (in steps of $\mathrm{0.05\ R_{200c}}$) and $\mathrm{\pm 0.5\ R_{200c}}$ along the minor axis (in steps of $\mathrm{0.05\ R_{200c}}$). If we assume that each edge-on disk galaxy is approximately symmetric along both axes, there are 4 pointings per major-minor axis position in each galaxy. For each galaxy, there are 651 sight-lines and therefore 55,986 across our MW analog sample. We ignore two subhalos in our calculation (513845, 527309), since at least one sight-line in each galaxy did not contain gas. These subhalos are missing gas at large radii and we don't expect their exclusion to significantly change our results.     

For each sightline, we produce spectra for the following lines:
\noindent
\begin{center}
    Mg~\textsc{ii} 2796 \text{\AA} \ \ \ \     Si~\textsc{iv} 1403 \text{\AA}\\
    C~\textsc{iv} 1548 \text{\AA} \ \ \ \ \ O~\textsc{vi} 1038 \text{\AA}  \\
    
\end{center}

For each spectrum, we subtract off the bulk velocity of the halo and calculate the line-of-sight velocity component. Since each gas cell in the simulation is assigned an angular momentum misalignment angle, we also produce the component of the total absorption line that comes from aligned material. We compute the equivalent width (EW) of the total absorption line and corotating absorption line. We then compute the ratio (EW($X_i$, $\theta \leq 30^\circ$) / EW($X_i$, $\theta \leq 180^\circ$) (where $X_i$ refers to the ion of interest) for each ion as a function of major-minor axis position. To calculate the average angular momentum alignment fraction at a given location, we weight the value for each galaxy by the EW. In addition, we only include sightlines where the given absorber has an $\mathrm{EW> 30\,m}$\AA. This threshold was chosen based on typical low-redshift COS G160M EW detection limits with $\mathrm{S/N\sim10}$. This choice ensures that our sample will have minimal contribution from unobservable gas; note that reasonable variations in this threshold have very little impact on our results. 

Several observational studies define co-rotation as whether or not the absorption has the same Doppler shift sign as the disk rotation. In addition to computing the fraction of angular momentum aligned gas, we aim to make a more straightforward comparison to the observational definition. We calculate the fraction of the absorption with the the same Doppler shift sign. This is the same definition used in \cite{nateghi24} (see their Figure 2) and \cite{Nateghi2024_2,Kacprzak2025}. \cite{nateghi24} call this the equivalent co-rotation fraction, $f_\mathrm{EWcorot}$. In our grid of sightlines, we calculate $f_\mathrm{EWcorot}$ for each ion as a function of distance along the midplane and z-height. We also weight $f_\mathrm{EWcorot}$ by the EW of each sightline and remove any sightlines where $\mathrm{EW< 30\,m}$\AA.

We also translate the $f_\mathrm{EWcorot}$ to a binary statement on whether a system is consistent with co-rotation ($f_\mathrm{EWcorot} > 0.50$ or $f_\mathrm{EWcorot} > 0.90$). This can be compared to previous work (e.g., \cite{Ho2017,Ho2025,Martin2019,Kacprzak2010,Kacprzak2019}), where a system is considered corotating if its absorption is predominantly on one side of the galaxy systemic velocity. However, we caution directly comparing our work with the observational work, since some of our absorption is undetectable in observational work. See \S\ref{subsubsec:magnesium} for discussion on how we calculate the co-rotation fraction taking into account the extent of corotating absorption and \lq observability' of the gas.

\section{Results}

We now show the results of applying our angular momentum misalignment definition to  our IllustrisTNG sample. We inspect the angular alignment ($\theta$) in a single galaxy and the entire sample (\S\ref{subsec:MW_misalignment}), using both the column density (\S\ref{subsec:column_density},\ref{subsec:column_density_angular_momentum}) and ion mass (\S\ref{subsec:ion_mass_distribution}). In addition, we show how our angular momentum alignment manifests in a mock HST/COS survey and how it compares to the line-of-sight co-rotation fraction (\S\ref{subsec:absorption_line_study}). 

\label{sec:results}


\subsection{Misalignment Angle in TNG50 MW Sample}\label{subsec:MW_misalignment}

\begin{figure*}[t]
\includegraphics[width=\linewidth]{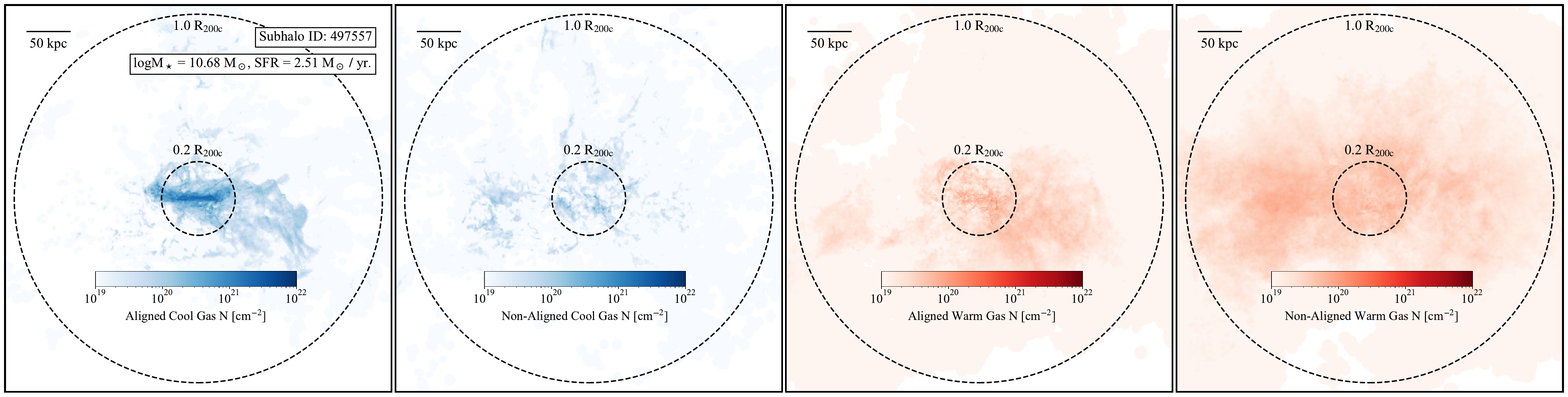}
\caption{For a single MW analog in our sample (Subhalo ID: 497557), the edge-on gas column density map of four distinct components: 1. the aligned cool ($10^4\ \mathrm{K} \leq \mathrm{T} < 10^5\  \mathrm{K}$) gas (far left), 2. the non-aligned cool gas (middle left), 3. the aligned warm ($10^5\  \mathrm{K} \leq \mathrm{T} < 10^6\  \mathrm{K}$) component (middle right), and 4. the non-aligned warm gas (far right). In each image, the inner circle is 0.20 $\mathrm{R_{200c}}$ and the outer circle is $\mathrm{R_{200c}}$.   
\label{fig:warmcool_image}}
\end{figure*}

The angular momentum misalignment angle distribution for one of the galaxies (Subhalo ID: 497557) is shown in Figure \ref{fig:warmcool_image}. This MW analog has a stellar mass ($\log \mathrm{M}_\ast = 10.68\ \mathrm{M_\odot}$) and star-formation rate ($\log \mathrm{SFR} = 0.40\ \mathrm{M_\odot\ yr^{-1}}$) and is in the approximate center of our sample selection in Figure \ref{fig:selection}. We highlight four distinct components of the galaxy: 1. the aligned cool gas (far left), 2. non-aligned cool gas (middle left), 3. aligned warm gas (middle right), and 4. non-aligned warm gas (far right). We separate the cool gas ($10^4\ \mathrm{K} \leq \mathrm{T} < 10^5\  \mathrm{K}$) and warm gas ($10^5\  \mathrm{K} \leq \mathrm{T} < 10^6\  \mathrm{K}$) to broadly distinguish between cool gas traced by Mg $\textsc{ii}$ and warm gas traced by O $\textsc{vi}$. The virial temperature of our MW analogs is typically in the upper part of the warm temperature range ($(5$--$10)\times10^5\,\mathrm{K}$). Si $\textsc{iv}$ and C $\textsc{iv}$ trace an approximate temperature range bridging Mg~\textsc{ii} and O~\textsc{vi} \citep{WerkReviewArticle}. There is no gas below $10^4\ \mathrm{K}$, since we exclude star-forming gas. 

In Figure \ref{fig:warmcool_image}, there is a clear distinction in the spatial gas distribution. There is a substantial fraction of the cool gas that is aligned and is located either near the star-forming disk or in a filament or cloud-like feature that is accreting onto the disk. In this example subhalo,  the aligned cool gas spatial distribution is asymmetric, with more gas present on one side of the major axis. It is possible that this material is associated with a recent merger. This lopsided feature also demonstrates the issue with viewing the edge-on galaxy from one point-of-view. This is especially the case with absorption-line spectroscopy, where it is difficult to determine where along the line-of-sight the gas is located. With a different edge-on rotation in the simulation, the aligned cool gas may appear more uniform. 


In comparison, only a small fraction of the cool gas is non-aligned. The origin of this material is uncertain, though it may be associated with outflows along the minor axis, accretion onto the halo, or gas that has precipitated from the hot halo. Some aligned warm gas traces the aligned cool gas within the ISM; however, a substantial fraction of the aligned warm gas lies beyond the inner 0.20 $\mathrm{R_{200c}}$ (inner circle), preferentially along the major axis. By contrast, the non-aligned warm gas is more volume-filling and extends out to $\mathrm{R_{200c}}$ (outer circle), with a larger fraction of the gas residing along the major axis.

Although a detailed study of gas accretion in our sample is beyond the scope of this work, we briefly comment on the relationship between angular momentum misalignment and radial velocity. Similar to Figure \ref{fig:warmcool_image}, we examine aligned and misaligned gas as a function of phase and location within either the ISM $\mathrm{(r/R_{200c} < 0.2)}$ or CGM $\mathrm{(r/R_{200c} \geq 0.2)}$. For each phase, alignment, and location, we compute the ratio of gas mass that is inflowing $(v_\mathrm{r} < 0)$ to gas mass that is outflowing $(v_\mathrm{r} \geq 0)$. We find that a typical MW analog in our sample has $\sim2\times$ more mass inflowing than outflowing. The only exception is warm gas (both aligned and misaligned) within the ISM, where mass outflow dominates, likely driven by feedback processes.

We also calculate the net mass flow rate and mass-weighted net radial velocity as functions of radius and azimuthal angle relative to the major and minor axes \citep[see Figure 2 in][]{Peroux2020}. We find that the net mass flow rate along the major axis (within $\sim30^\circ$ of the major axis) is dominated by inflow out to $\sim\mathrm{R_{200c}}$. When separated by temperature, cold ($\sim10^4$ K) gas dominates the inflow along the major axis. Warm gas ($\sim10^{5}$--$10^{6}$ K) also exhibits a net inward mass flow beyond $\sim0.1$--$0.2\,\mathrm{R_{200c}}$, though the cold phase remains the dominant contributor to the mass flux. In contrast, the net mass flow rate along the minor axis is dominated by outflow at all temperatures.

The mass-weighted net radial velocity reveals a similar picture of accretion along the midplane. Both cool and warm gas exhibit inward radial velocities that stall at $\sim0.1$--$0.2\,\mathrm{R_{200c}}$. This radius approximately corresponds to $\mathrm{R_{DLA}}$, defined as the characteristic radius where the neutral hydrogen column density drops below $10^{20.3}\,\mathrm{cm^{-2}}$. This likely marks the outer disk region where gas decelerates as it approaches the star-forming disk, consistent with the behavior reported in \citet{Trapp2022}. Along the minor axis, cool gas is primarily outflowing but slows to $v_\mathrm{r} \sim 0\ \mathrm{km\,s^{-1}}$ near the disk, consistent with gas recycling, while the warm phase remains dominated by outward radial motion.

Taken together, these results indicate that the majority of the aligned gas we identify is inflowing, predominantly along the major axis. A more detailed analysis of the mass accretion history of the MW TNG50 sample and the disk--halo interface will be presented in future work (M. Messere et al. 2026, in preparation).

\begin{figure*}
\includegraphics[width=\linewidth]{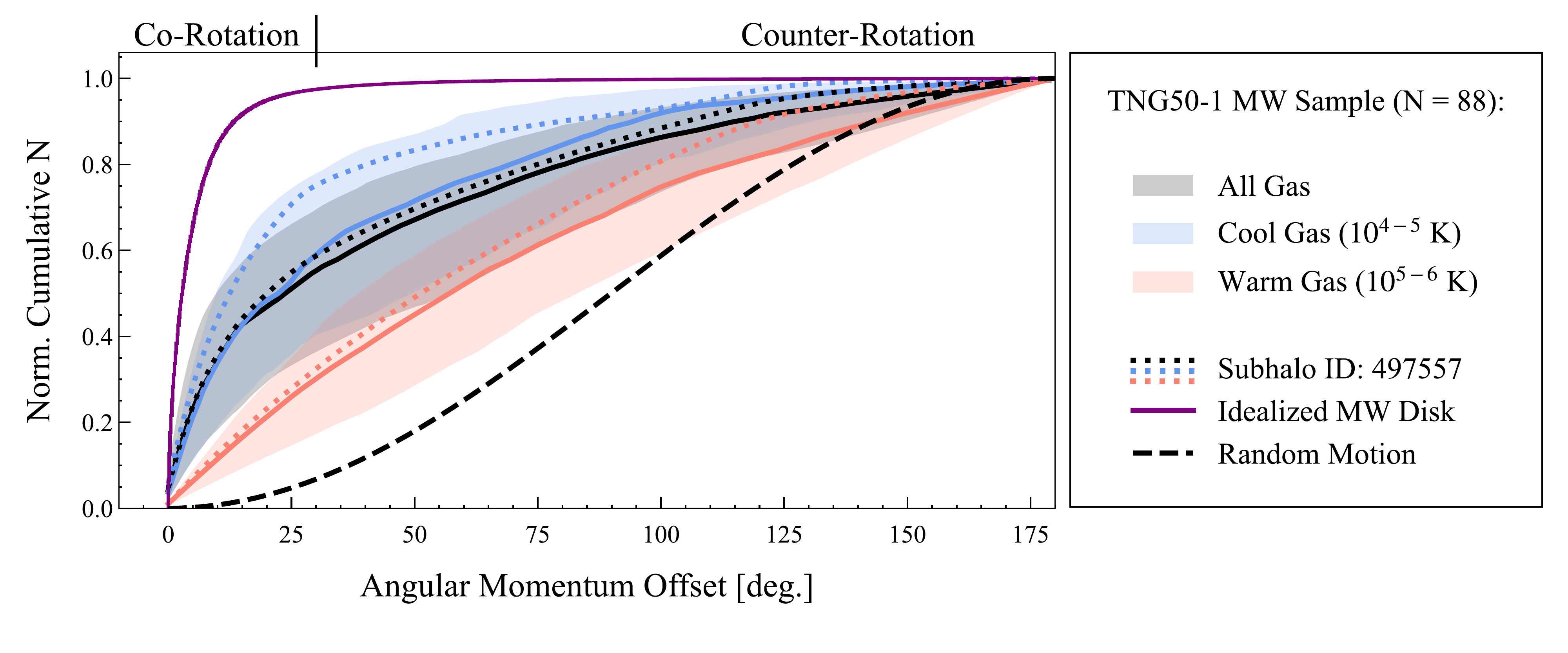}
\caption{The median normalized cumulative distribution of the angular momentum misalignment angle in the TNG50 MW analog sample for all gas (solid black), cool gas (solid blue), and warm gas (solid red). We include all gas within each subhalo. The shaded regions are bounded by the 16th percentile and 84th percentile of their respective distribution. This is compared to the distribution in the idealized disk galaxy (\S\ref{subsec:agora}) (solid purple), where nearly all of the gas is aligned. 
We also include the distribution of the individual MW analog in Figure \ref{fig:warmcool_image} (Subhalo ID: 497557) along with the distribution from random motion (\S\ref{subsec:define misalignment angle}).      
\label{fig:cumulative_TNG}}
\end{figure*}

In Figure \ref{fig:cumulative_TNG}, we show the median cumulative distribution of the angular momentum offset for the entire MW sample as the solid black line. We include all gas within each subhalo. The shaded black region encloses the 16th percentile and 84th percentile. The same is shown for the cool gas (solid blue) and warm gas (solid red). We also show the cumulative distribution for the idealized MW disk (solid purple, \S\ref{subsec:agora}), the cumulative distribution from random motion in the halo (dashed black, \S\ref{subsec:define misalignment angle}), and the cumulative distribution for the halo in Figure \ref{fig:warmcool_image} (dotted). We note that Figure \ref{fig:cumulative_TNG} is the cumulative gas cell \emph{count} and does not consider the gas cell volume or mass. However, gas cells have a similar mass in IllustrisTNG. We explore the mass distribution in more detail in \S\ref{subsec:ion_mass_distribution}.   

In comparison to the idealized disk galaxy simulation, there is a broader distribution of the angular momentum misalignment angle in IllustrisTNG. This is a direct consequence of the cosmological environment in which our MW analogs evolve. There is the inflow/outflow of material, interaction with satellite systems, AGN feedback, the halo accretion shock, and several other factors that are not captured in an idealized disk simulation \citep{Fielding2020a}. The warm gas motion is less aligned with the galaxy angular momentum axis. This can be seen by the sharp rise in the cool gas cumulative distribution within $\theta\sim 30^\circ$. 60\% of the cool gas cell count is corotating whereas only 30\% of the warm gas cell count is corotating. There is a clear preference for the warm gas to have a misalignment angle $\theta \leq 30^\circ$, but there is a larger tail in the distribution (toward larger $\theta$) compared to the cool gas distribution. We also find that there is no MW analog that can be explained by the random motion of gas alone.   

Lastly, we find that there are a few exceptions\footnote{Subhalo IDs: 469487, 482889, 561676, 514829, 513845, 510585, 517271, 528322, 580406, and 552414} to the sample median in Figure \ref{fig:cumulative_TNG}, with such galaxies accounting for $\sim 10\%$ of the sample. This is most clear in Subhalo ID 469487, where the CGM line-of-sight velocity is flipped along the major axis beyond $\gtrsim 0.20\ \mathrm{R_{200c}}$. In addition, Subhalo ID 510585 co-rotation is symmetric about the major axis, not the minor axis.   

\subsection{Column Density}\label{subsec:column_density}

\begin{figure}
\centering
\includegraphics[height=0.95\textheight]{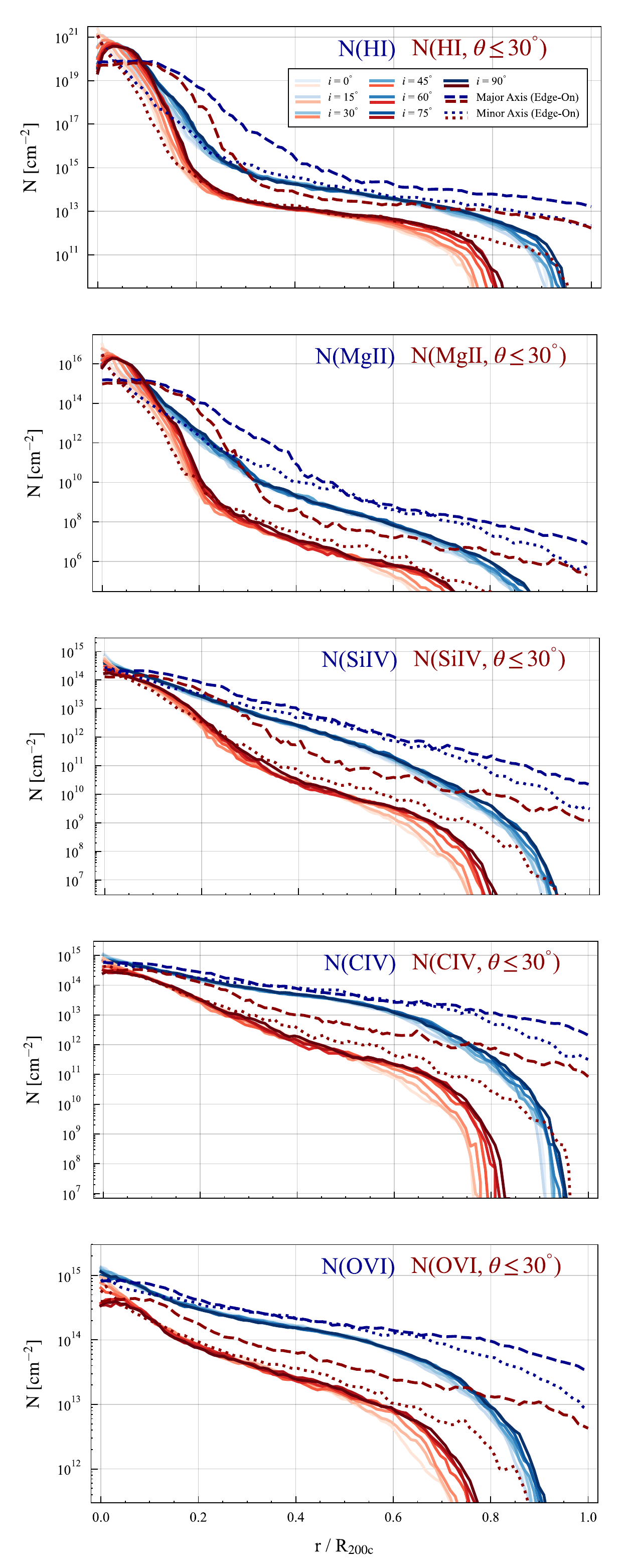}
\caption{The median column density profile for H$\textsc{i}$, Mg$\textsc{ii}$, Si$\textsc{iv}$, C$\textsc{iv}$, and O$\textsc{vi}$ (top to bottom). The total ion column density (blue) and corotating ion column density (red) are shown for the major (dashed) and minor (dotted) axes. The radial column density profile is also calculated as a function of inclination angle (light-to-dark blue and red).  
\label{fig:column_density}}
\end{figure}

We now highlight the column density profile and corresponding angular momentum alignment sub-component ($\theta <30^\circ$) in Figure \ref{fig:column_density}. See \S\ref{subsec:column_density_corotation_fraction} on how this is calculated. Here, we present the median column density profile as a function of 1. ion, 2. angular momentum alignment, 3. axis, and 4. inclination. The total ion column density ($\theta \leq 180^\circ$) profile for the major axis (dashed blue) and minor axis (dotted blue) demonstrate that the higher the ionization state, the more axis-symmetric (volume-filling) the column density distribution. For example,  N(H~\textsc{i}) along the major axis is larger than the minor axis at nearly all distances. $\log\,$N(H~\textsc{i}) $\sim{20}$ cm$^{-2}$ out to $0.1\ \mathrm{R_{200c}}$ along the major axis and gradually declines to $\log\,$N(H~\textsc{i}) $\sim14-15$ cm$^{-2}$  at $0.5\ \mathrm{R_{200c}}$. In comparison, N(H~\textsc{i}) along the minor axis steeply declines outside of the immediate disk. Beyond $0.5\ \mathrm{R_{200c}}$, N(H~\textsc{i}) along the minor axis is typically 10\% of the N(H~\textsc{i}) value along the major axis. A similar picture is seen in Mg~\textsc{ii}. In the disk-halo interface region ($0.05 \lesssim \mathrm{r/R_{200c}} \lesssim 0.20$), N(Mg~\textsc{ii}) similarly dominates along the major axis compared to the minor axis.  $\log\,$N(Mg~\textsc{ii}) $\sim15$ cm$^{-2}$  out to $0.1\ \mathrm{R_{200c}}$ along the major axis and gradually declines to $\log\,$N(Mg~\textsc{ii}) $\sim10$ cm$^{-2}$ at $0.5\ \mathrm{R_{200c}}$.  However, beyond $0.5\ \mathrm{R_{200c}}$, N(Mg~\textsc{ii}) is nearly the same, independent of which axis it resides on.   

For the warmer ions (Si~\textsc{iv}, C~\textsc{iv}, O~\textsc{vi}), there is less azimuthal dependence of the column density, with only a slight enhancement $\mathrm{\lesssim 0.5\ R_{200c}}$. In addition, the column density profile is more shallow compared to H~\textsc{i} and Mg~\textsc{ii}. N(H~\textsc{i}) and N(Mg~\textsc{ii}) along the major axis decline by $\sim10^5$ in the inner $\mathrm{0.5\ R_{200c}}$. In comparison, N(Si~\textsc{iv}), N(C~\textsc{iv}), and N(O~\textsc{vi}) only decline by $\sim10$ in the same region. $\log\,$N(Si~\textsc{iv}), $\log\,$N(C~\textsc{iv}), and $\log\,$N(O~\textsc{vi}) are $\sim{14},\ {14.5},\ {15}$ cm$^{-2}$  at $0.1\ \mathrm{R_{200c}}$, respectively. 

There is no clear dependence on the inclination angle. In H~\textsc{i} and Mg~\textsc{ii}, there is a slight enhancement in the column density $\mathrm{\lesssim0.2\ R_{200c}}$, owing to the extent of the disk. The edge-on column density is $\sim10\%$ higher than the face-on column density. This is not the case for the warmer ions, since they are more uniformly distributed throughout the galaxy. 

We also note that the effect of using only the subhalo cutout for each galaxy can be seen at large radii, where there is a nonphysical turnover in the column density value. This occurs at about 0.8 $\mathrm{R_{200c}}$ for the column density profile that includes all gas ($\theta \leq 180^\circ$) and about 0.75 $\mathrm{R_{200c}}$ for the column density profile that includes only aligned gas ($\theta \leq 30^\circ$).

The column density profile of the aligned gas has a different behavior than the total column density profile. In general, the aligned column density profile drops off more steeply than the corresponding total column density profile. This is a natural consequence of more aligned material found closer to the star-forming disk. We also find a turnover radius in the aligned major axis profile at $\mathrm{\sim 0.3\ R_{200c}}$ that is not seen in the total column density profile. This is most clear in H~\textsc{i} and Mg~\textsc{ii}.   

In H~\textsc{i} and Mg~\textsc{ii}, the aligned column density nearly traces the total column density $\mathrm{\lesssim 0.1\ R_{200c}}$. Nearly all observed N(H~\textsc{i}) and N(Mg~\textsc{ii}) within $\mathrm{\lesssim 0.1\ R_{200c}}$ is aligned with the star-forming disk, likely owing to our angular momentum axis defined using gas in the inner 30 kpc ($\mathrm{\sim 0.1\ R_{200c}}$). This is less clear in the warmer ions, since there is more likely to be warm non-aligned material in the outskirts of the halo that contribute to the column density. This is also seen in Figure \ref{fig:warmcool_image}. 

\subsection{Angular Momentum Alignment Fraction}\label{subsec:column_density_angular_momentum}

\begin{figure}
\centering
\includegraphics[height=0.85\textheight]{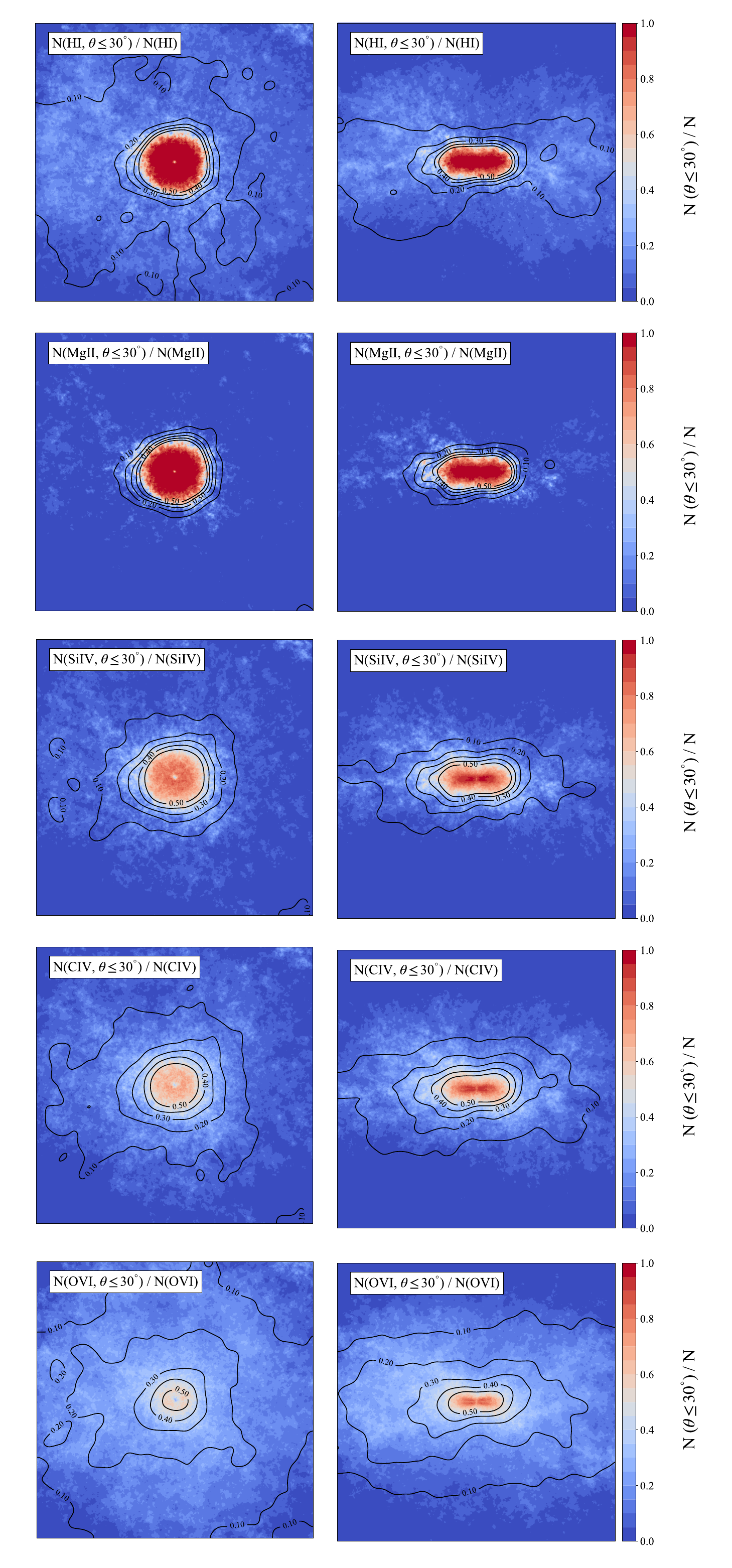}
\caption{For each ion, the $i=0^\circ$ (left) and $i=90^\circ$ (right) median fraction of the total column density from angular momentum aligned gas. Each image is 1 $\mathrm{R_{200c}} \times 1 \mathrm{R_{200c}}$. In the edge-on view, it is clear that each ion is aligned in a disklike structure, even for warmer gas (e.g., O~\textsc{vi}).}    
\label{fig:corotation_fraction}
\end{figure}

In order to better visualize the structure of the angular momentum alignment sub-component, we show in Figure \ref{fig:corotation_fraction} the fraction of the total column density (of each line-of-sight) that is aligned. We show the sample median angular momentum alignment fraction for each ion and inclination ($0^\circ,\, 90^\circ$). In the face-on perspective (left), nearly 100 percent of the observed H\textsc{i} and Mg~\textsc{ii} in the inner $0.2\ \mathrm{R_{200c}}$ is aligned. Similar to what is shown in Figure \ref{fig:column_density}, a smaller fraction of the higher ionization gas in the inner disk is aligned. At $\sim 0.5\ \mathrm{R_{200c}}$, $\sim 20$\% of the O~\textsc{vi} column density is aligned.

For the lower ionization species, the column density distribution shifts from circular to disklike, since this gas is largely found within the (extended) disk. For the higher ionization species, the column density distribution remains nearly circular, independent of inclination. N(O~\textsc{vi}) in the edge-on case remains uniformly distributed throughout the CGM. \emph{All} of the ions become more disklike in angular momentum alignment as the inclination shifts from face-on to edge-on. Despite O~\textsc{vi} having a nearly symmetric column density profile along the major axis and minor axis, there is a clear subset that is aligned preferentially along the major axis compared to the minor axis. 

In the typical IllustrisTNG MW, at about $\mathrm{0.5\ R_{200c}}$, 20-30\% of the observed N(O~\textsc{vi}) along the major axis is from aligned gas. At the same distance, but along the minor axis, only 10 percent of the observed O~\textsc{vi} column density is from corotating gas. For H\textsc{i}, this is 10-20\% along the major axis and less than 10\% along the minor axis. In Figure \ref{fig:corotation_fraction}, there is also some flaring of the disklike structures seen in the median H\textsc{i} stack, where the 10\% contour has a flare-like structure. Aligned Mg~\textsc{ii} is only found near the inner disk. Si~\textsc{iv} and C~\textsc{iv} follow nearly the same behavior as H\textsc{i}, but there is a slight enhancement in the amount of aligned gas along the minor axis.

We report the semi-minor axis (b) to semi-major axis (a) ratio for each ion and inclination angle in Table \ref{table:ba_ratio}. For each ion and inclination, we identify the b/a ratio where the angular momentum alignment column density fraction is 20\% and 50\%. The first value reported is the b/a ratio and the next two are the semi-minor and semi-major axes. These are given as a fraction of $\mathrm{R_{200c}}$. Since the distribution may be somewhat asymmetric on either side of the major and minor axis, we calculate the average of the semi-major and semi-minor axes. 

Table \ref{table:ba_ratio} highlights that in the edge-on case, the higher the ionization state, the less disklike the aligned column density. For example, when $\sim50\%$ of  N(Mg~\textsc{ii}) is aligned with the angular momentum of the galaxy, the edge-on galaxy sample has a b/a ratio of 0.49 (b = 0.11, a = 0.23). Therefore, N(Mg~\textsc{ii}) that is $\gtrsim50\%$ aligned with the angular momentum of the galaxy is preferentially found along the midplane (out to 0.23 $\mathrm{R_{200c}}$) compared to the minor axis (out to 0.11 $\mathrm{R_{200c}}$). The edge-on N(Mg~\textsc{ii}) b/a ratio for $20\%$ angular momentum alignment is still $\sim0.5$ (b = 0.17, a = 0.37). This suggests that N(Mg~\textsc{ii}) is still disklike, even at a larger radii where there is less ($20\%$) material aligned with the angular momentum of the galaxy.

In comparison, aligned N(O~\textsc{vi}) has a more oblate-spheroid-like structure. The N(O~\textsc{vi}) b/a ratio is 0.56 (b = 0.07, a = 0.13) for 50\% angular momentum alignment, \emph{but} 0.33 (b = 0.26, a = 0.81) for 20\% angular momentum alignment. Therefore, there is $20\%$ angular momentum alignment out to nearly the halo boundary ($0.81\,\mathrm{R_{200c}}$) along the major axis but not the minor axis. This is also seen in Figure \ref{fig:corotation_fraction} (lower-right), where we have labeled the N(O\textsc{vi}) alignment $20\%$ contour. It is not clear if this O~\textsc{vi} structure in the CGM is a corotating thick disk or a combination of both a corotating disk component and corotating halo component. We discuss this further in \S\ref{subsubsec: Warm Gas}. 

\begin{deluxetable*}{cccccccc} 
\tablenum{1}
\tablecaption{The semi-minor axis (b) to semi-major axis (a) ratio for each ion and inclination angle in Figure \ref{fig:corotation_fraction}. The value of b and a are also given as a fraction of $\mathrm{R_{200c}}$. This is calculated where 20\% of the ion column density is from aligned material and where 50\% of the ion column density is from aligned material. For example, when $\sim$ 20\% of the total O~\textsc{vi} column density is from aligned O~\textsc{vi}, the face-on galaxy sample ($i$ = 0$^\circ$) has a b/a ratio of 0.81 (b = 0.45, a = 0.55). In comparison, the edge-on galaxy sample ($i$ = 90$^\circ$) has a b/a ratio of 0.33 (b = 0.26, a = 0.81). The aligned O~\textsc{vi} is preferentially located along the midplane out to very large radii ($\sim $ 0.81 $\mathrm{R_{200c}}$), even if it may not dominate the total column density.}
\label{table:ba_ratio}
\tablehead{\colhead{$i$ [deg.]} & \colhead{H\textsc{i}} & \colhead{Mg~\textsc{ii}}  & \colhead{Si~\textsc{iv}} & \colhead{C~\textsc{iv}} & \colhead{O~\textsc{vi}}}
\startdata
0 (20\%) & 0.82 (0.30,0.37) & 0.87 (0.27,0.32) & 0.81 (0.31,0.38) & 0.82 (0.32,0.38) & 0.81 (0.45,0.55) \\
0 (50\%) & 0.93 (0.22,0.24) & 0.93 (0.22,0.23) & 0.93 (0.19,0.21) & 0.92 (0.15,0.16) & 0.78 (0.05,0.06)  \\ \\
45 (20\%) & 0.47 (0.26,0.55) & 0.68 (0.23,0.34) & 0.64 (0.26,0.41) & 0.60 (0.28,0.46) &  0.47 (0.33,0.71) \\
45 (50\%) & 0.70 (0.18,0.25) & 0.67 (0.17,0.25) & 0.66 (0.15,0.23) & 0.62 (0.12,0.20) & 0.66 (0.07,0.11) \\ \\
90 (20\%) & 0.39 (0.18,0.45) & 0.45 (0.17,0.37) & 0.47 (0.20,0.43) & 0.44 (0.22,0.51) & 0.33 (0.26,0.81) \\
90 (50\%) & 0.50 (0.12,0.23) & 0.49 (0.11,0.23) & 0.49 (0.11,0.22) & 0.59 (0.10,0.17) & 0.56 (0.07,0.13) 
\enddata
\end{deluxetable*}

\subsection{Ion Mass Distribution}\label{subsec:ion_mass_distribution}

\begin{figure*}
\centering
\includegraphics[width=1\linewidth]{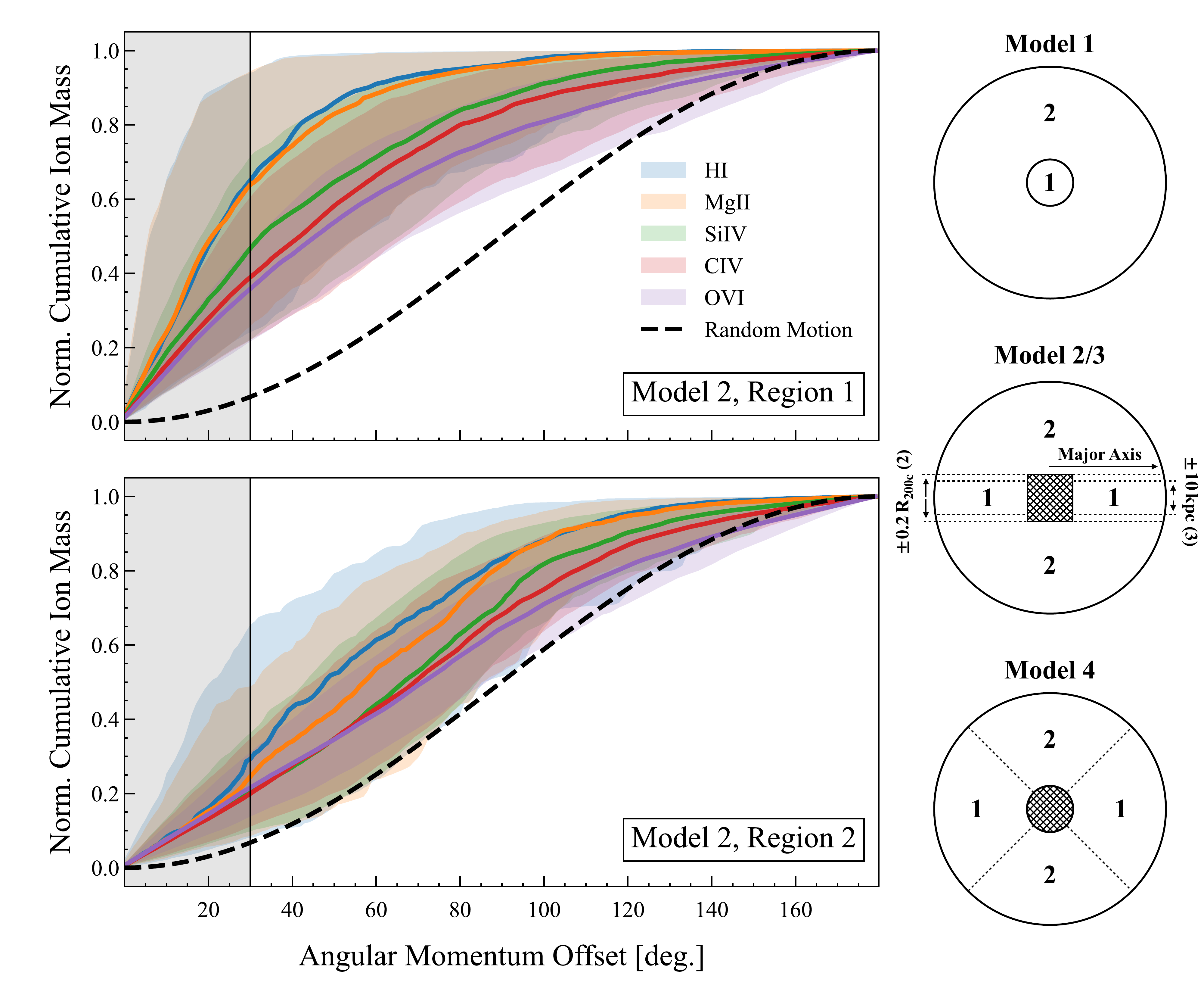}
\caption{The normalized cumulative ion mass distribution as a function of angular momentum alignment angle in the Model 2, Region 1 (top panel) and Model 2, Region 2 (lower panel). For each ion, the solid line represents the median cumulative distribution and the shaded region encapsulates the 16th and 84th percentiles. We also include the upper limit of our angular momentum alignment definition (solid black; $\mathrm{\theta = 30^\circ}$) and the cumulative distribution from a random $\theta$ distribution (dashed black). We depict our four models on the right, where the major axis is horizontal. 
\label{fig:ions}}
\end{figure*}

In this section, we quantify the relationship between the ion mass (\S\ref{subsec:trident}) and the angular momentum offset of the gas (Figure~\ref{fig:ions}).  We compute this for four different models, where each model is a different way of spatially dividing the ISM and CGM into 2 regions: 1. the ISM (r/$\mathrm{R_{200c}}$ $< 0.2$) and CGM (r/$\mathrm{R_{200c}}$ $\geq 0.2$), 2. within an extended thick disk ($\mathrm{z/R_{200c}}<0.2$) and outside an extended thick disk ($\mathrm{z/R_{200c}}\geq0.2$), 3. within an extended thin disk ($\mathrm{z/R_{200c}}<10\, \mathrm{kpc}$) and outside an extended thin disk ($\mathrm{z/R_{200c}}\geq10\, \mathrm{kpc}$), and 4. within a wedge-like region along the major axis ($\phi\leq 45^\circ$) and within a wedge-like region along the minor axis ($\phi>45^\circ$), where $\phi$ is the azimuthal angle measured from the major axis. The ISM is excluded in the last three models. The four different models are depicted in Figure~\ref{fig:ions}. 

The boundary between the ISM and CGM is chosen from Figure \ref{fig:corotation_fraction}, where the bulk of the total column density is found within 0.2 $\mathrm{R_{200c}}$. In addition, we choose Model 3 based on work measuring the extent of C~\textsc{iv} in the MW (\cite{Bish2021}) and Model 4 based on observational work that separates major axis versus minor axis absorption (e.g., \cite{Bordoloi2011}). Model 2 is a thick disk model, between Model 3 and Model 4. For each model and region, we calculate the cumulative mass distribution of each ion contained within a given misalignment angle range and normalize the result by the total ion mass in each region. For example, we display the Model 2 cumulative mass distribution in Figure~\ref{fig:ions} for Region 1 (top) and Region 2 (bottom). For each ion, we include the median as the solid line and the 16th and 84th percentiles as the shaded region. In addition, we include the expectation from random gas motion and a vertical line representing the upper limit of our angular momentum alignment definition ($\theta=30^\circ$).

As shown in Figure~\ref{fig:ions}, there is a clear distinction between the fraction of aligned mass along the major axis (Region 1) compared to the minor axis (Region 2). For example, the typical MW analog has $\sim 65\%$ ($\sim25\%$) of the total H~\textsc{i} and Mg~\textsc{ii} mass aligned along the major (minor) axis. The median angular momentum alignment mass fraction is 45\% (19\%), 38\% (18\%), and 35\% (20\%) along the major (minor) axis for Si~\textsc{iv}, C~\textsc{iv}, and O~\textsc{vi}, respectively. The 16th percentile for H~\textsc{i} and Mg~\textsc{ii} is $\sim25\%$. This is likely driven by halos with a large fraction of their CGM counter-rotating. On the contrary, there are also some halos that have a substantial amount of aligned mass. The 84th percentile for H~\textsc{i} and Mg~\textsc{ii} is $94\%$. We note that this spread becomes much smaller as the ionization energy increases. The difference between the 16th and 84th percentiles are 49\%, 38\%, and 25\% for Si~\textsc{iv}, C~\textsc{iv}, and O~\textsc{vi}, respectively. We also find that the median Region 1 and Region 2 cumulative mass distribution for each ion is inconsistent with random motion -- there is still some level of angular momentum alignment, even for the warmer gas traced by O~\textsc{vi}. 

For a more complete census of the angular momentum alignment fraction in each model and region, we include our results in Table \ref{table:corotation_fractions}. For each model, the top row corresponds to Region 1 and the bottom row corresponds to Region 2. The 16th, 50th, and 84th percentiles and the median total mass ($\mathrm{\log_{10}(M/M_\odot)}$) are given for each ion and region. We include the same results for the entire halo in the bottom row. In Model 1, $\sim2\times$ more H~\textsc{i} and Mg~\textsc{ii} mass is corotating in the ISM (Region 1) compared to the CGM (Region 2). In comparison, this factor is $\sim3$ for Si~\textsc{iv}, C~\textsc{iv}, and O~\textsc{vi}. 

We also find that Model 3 and Model 4 are at the extreme ends of Model 2. For example, the alignment fraction in Model 3 along the major axis \emph{and} minor axis is higher than Model 2. This suggests that there is a large fraction of angular momentum alignment directly along the plane ($\mathrm{|z|\pm10\ kpc}$). In comparison, Model 4, Region 1 is slightly lower than Model 2, Region 1, suggesting the wedge-like region is being diluted by non-aligned material.  

\begin{deluxetable*}{cccccccc} 
\tablenum{2}
\tablecaption{The aligned fraction ($\theta\leq 30^\circ$) for each model, region, and ion. The first row for each column corresponds to Region 1 and the second row for each model corresponds to Region 2. The models are depicted in Figure \ref{fig:ions}. For each ion, the 16th, 50th, and 84th percentiles (over the sample of galaxies) of the aligned mass fraction are shown, followed by the total ion mass in the region ($\mathrm{log_{10}(M_{\it{X_i}}/M_\odot})$). In addition, we display the total subhalo angular momentum alignment fraction in the bottom row, followed by the total ion mass present in the halo. For example, in Model 2, Region 1, the median fraction of aligned O~\textsc{vi} mass is 35\%, where this region is $\sim10\%$ of the halo O~\textsc{vi} mass budget.}   
\label{table:corotation_fractions}
\tablehead{\colhead{Model/Region}  &  \colhead{H\textsc{i}} & \colhead{Mg~\textsc{ii}}  & \colhead{Si~\textsc{iv}} & \colhead{C~\textsc{iv}} & \colhead{O~\textsc{vi}}}
\startdata
1/1 & (0.85,0.97,0.99) (10.0) & (0.89,0.98,0.99) (5.4) & (0.58,0.78,0.88) (3.7) & (0.55,0.73,0.84) (4.1) & (0.48,0.65,0.76) (4.2) \\
1/2  & (0.19,0.59,0.93) (9.0) & (0.19,0.55,0.91) (4.2)  & (0.16,0.29,0.55) (4.1) & (0.15,0.25,0.41) (5.0) & (0.17,0.24,0.35) (5.3) \\ \\
2/1  &  (0.24,0.64,0.94) (8.8) & (0.25,0.63,0.94) (4.0) & (0.21,0.45,0.70) (3.7) & (0.21,0.38,0.59) (4.4) & (0.21,0.35,0.46) (4.8) \\
2/2  & (0.08,0.26,0.56) (8.0) & (0.08,0.24,0.47) (3.3)  & (0.10,0.19,0.34) (3.9) & (0.10,0.18,0.31) (4.8) & (0.13,0.20,0.29) (5.2) \\ \\
3/1  & (0.24,0.70,0.98) (8.5) & (0.25,0.70,0.98) (3.6) & (0.27,0.56,0.80) (3.2) & (0.23,0.44,0.69) (3.8)& (0.21,0.39,0.53) (4.2)\\
3/2  & (0.28,0.66,0.86) (9.0) & (0.32,0.60,0.83) (4.3)  & (0.18,0.30,0.54) (4.1) & (0.16,0.25,0.40) (5.0) & (0.17,0.25,0.35) (5.3) \\ \\
4/1  & (0.22,0.61,0.93) (9.0) & (0.22,0.58,0.93) (4.1) & (0.16,0.32,0.58) (4.0) & (0.15,0.25,0.45) (4.8) & (0.18,0.27,0.39) (5.2) \\
4/2  & (0.11,0.35,0.63) (8.0) & (0.11,0.31,0.58) (3.2)  & (0.11,0.23,0.39) (3.6) & (0.12,0.21,0.35) (4.5) & (0.12,0.20,0.29) (4.8) \\ \\
Entire Subhalo & (0.76,0.93,0.98) (10.0) & (0.82,0.95,0.99) (5.5) & (0.27,0.42,0.63) (4.3) & (0.20,0.32,0.46) (5.0) & (0.20,0.29,0.38) (5.4) \\
\enddata
\end{deluxetable*}

\subsection{Absorption Line Study}\label{subsec:absorption_line_study}

We have now shown the extent of angular momentum aligned gas in IllustrisTNG MW analogs in several distinct ways. In order to better connect to observational work, we produce mock absorption-line sightlines (\S\ref{subsec:sightline_generation}) and calculate (1) the alignment fraction along the line-of-sight and (2) the EW-weighted co-rotation fraction ($f_\mathrm{EWcorot}$). 

\subsubsection{Angular Momentum Alignment Angle}\label{subsubsec:angular_momentum}

\begin{figure}
\centering
\includegraphics[width=0.48\textwidth]{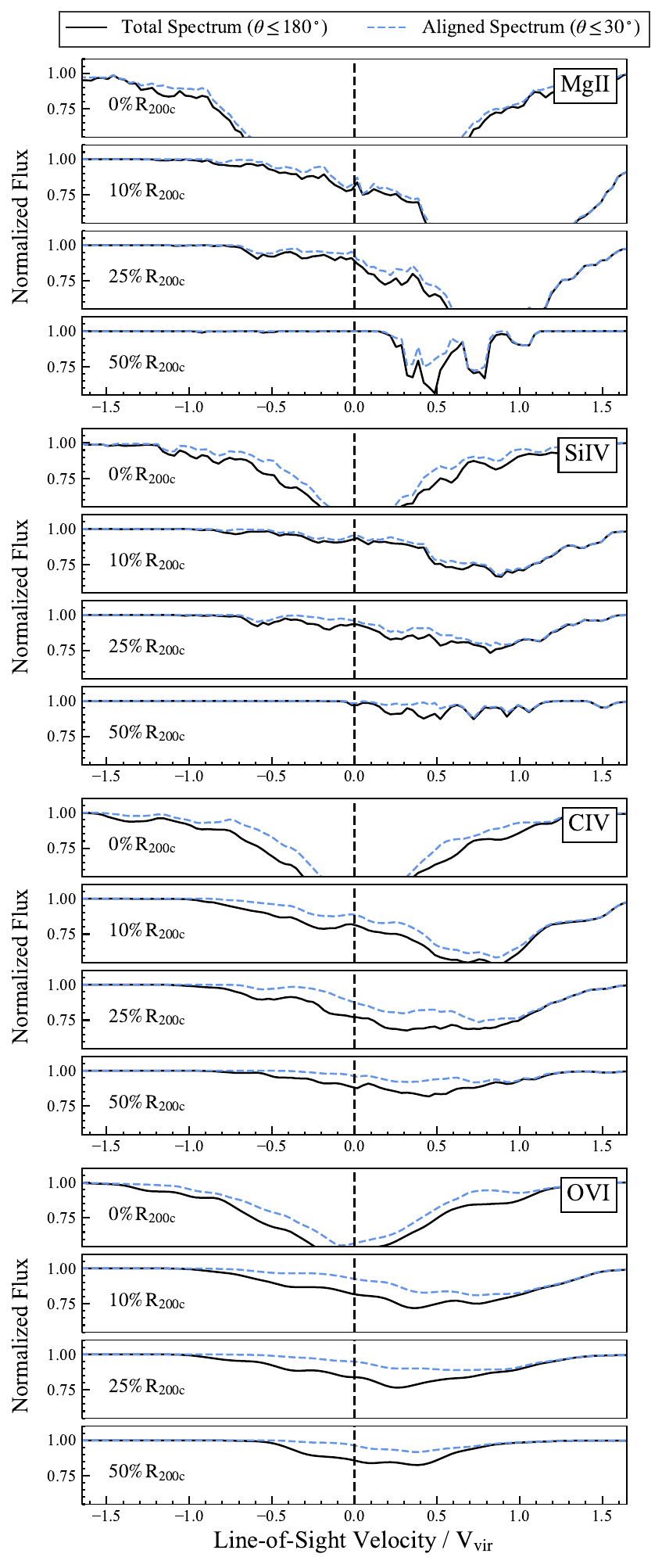}
\caption{The average spectra at $\mathrm{0\%\, R_{200c}}$, $\mathrm{10\%\, R_{200c}}$, $\mathrm{25\%\, R_{200c}}$, and $\mathrm{50\%\, R_{200c}}$ for the all gas (black) and aligned gas (blue). We find that although the absorption at large radii is weak, it is typically aligned and corotating with the galaxy. This is especially true for lower ions (e.g.,  Mg~\textsc{ii}).       
\label{fig:spectra}}
\end{figure}

In Figure \ref{fig:spectra}, we show the average spectrum for each ion at four different distances along the midplane. For each spectrum, we include all gas (solid black) and the aligned component (dashed blue). At $0\%\, \mathrm{R_{200c}}$ (the center of the galaxy), all ions are aligned with the systemic velocity of the galaxy, as expected ($\sim0\,\mathrm{km\,s^{-1}}$). In the inner CGM, the total spectrum is composed almost entirely of aligned gas. This decreases as the distance along the midplane and ionization increases. At $10\%\, \mathrm{R_{200c}}$, all of the ions trace a similar velocity ($\sim0.9\,\mathrm{V_{vir}}$). This is also true at $25\%\, \mathrm{R_{200c}}$ and $50\%\, \mathrm{R_{200c}}$, but the Mg~\textsc{ii} spectra become more noisy. By $50\%\, \mathrm{R_{200c}}$, the ions (if detectable) trace gas that is moving $\sim0.25-0.5\,\mathrm{V_{vir}}$ along the line-of-sight.           

\begin{figure*}
\centering
\includegraphics[width=\linewidth]{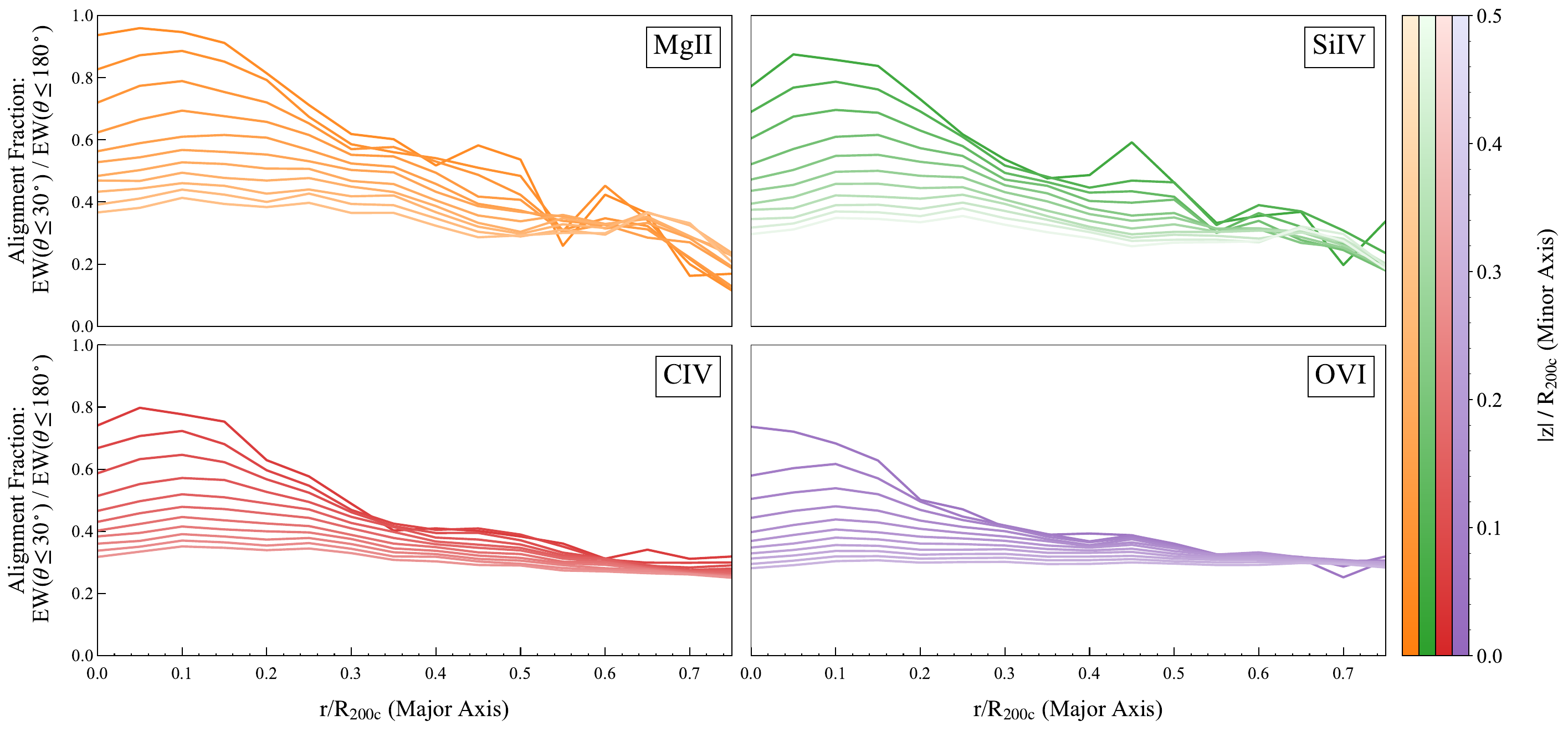}
\caption{The EW-weighted fraction of the total ion absorption from angular momentum aligned ($\theta \leq 30^\circ$) gas. We calculate this fraction as a function of major axis and minor axis and only include spectra where $\mathrm{EW> 30\,m}$\AA. We expect only $\sim10\%$ of gas to be corotating if the halo motion is random (see \S\ref{subsec:define misalignment angle}).           
\label{fig:absorption_fraction}}
\end{figure*}

To further investigate this behavior for each galaxy, sightline, and ion, the ratio of the aligned EW to the total EW (EW($X_i$, $\theta \leq 30^\circ$) / EW($X_i$, $\theta \leq 180^\circ$) is computed (\S\ref{subsec:sightline_generation}). In Figure \ref{fig:absorption_fraction}, we report the average angular momentum alignment fraction for each ion as a function of distance along the midplane and z-height. We find that nearly $80\%$ of the Mg~\textsc{ii} is aligned within $\mathrm{0.2\ R_{200c}}$ along the midplane. As the height above the midplane increases, the fraction of aligned gas decreases. For example, at $0.5 \,\mathrm{R_{200c}}$ above the midplane, the fraction of aligned Mg~\textsc{ii} within $\mathrm{0.2\ R_{200c}}$ goes down to $\sim0.4$. Beyond the inner CGM, the fraction of aligned Mg~\textsc{ii} decreases significantly. For the typical Mg~\textsc{ii} sightline beyond $\mathrm{0.5\ R_{200c}}$ at all z-heights, less than $50\%$ of the observable Mg~\textsc{ii} is aligned. The picture is similar for Si~\textsc{iv}, C~\textsc{iv}, and O~\textsc{vi}. The primary difference is that the aligned fraction within the inner $\mathrm{0.2\ R_{200c}}$ is lowered. Beyond $\mathrm{0.5\ R_{200c}}$, around $30\%$ of the O~\textsc{vi} is aligned with the angular momentum axis of the galaxy.

We also quantify the role of random gas motion. For every sightline, we re-run \texttt{TRIDENT} but assign a random misalignment angle to every gas cell along the line-of-sight. We find that only $\sim10\%$  of the alignment fraction is from random motion. Our results in Figure \ref{fig:absorption_fraction} cannot be explained by random motion in the CGM alone.    

\subsubsection{Observationally-Motivated Co-Rotation Fraction}\label{subsubsec:observationally_motivated}

\begin{figure*}
\centering
\includegraphics[width=\linewidth]{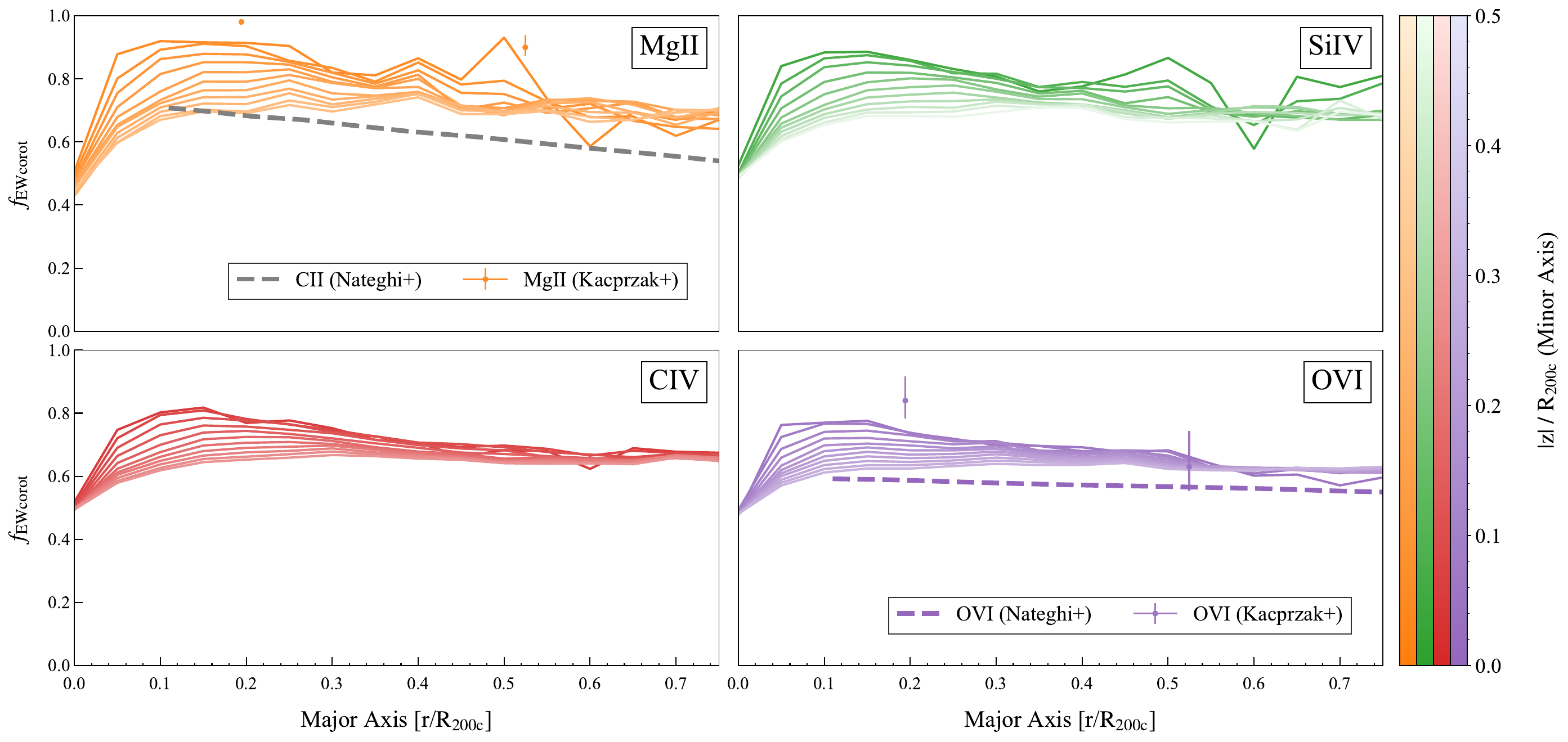}
\caption{$f_\mathrm{EWcorot}$ as a function of major axis and minor axis. We only include spectra where $\mathrm{EW> 30\,m}$\AA\, and use the EW as a weight in our average calculation at each sightline position. We find general agreement with \cite{Nateghi2024_2} and \cite{Kacprzak2025}. We use the average $\mathrm{R_{vir}}$ (129 kpc) in  \cite{Kacprzak2025} to scale their average inner and outer CGM distance.         
\label{fig:Nateghi_comparison}}
\end{figure*}

We present our $f_\mathrm{EWcorot}$ calculation for each ion in Figure \ref{fig:Nateghi_comparison}. We include the $f_\mathrm{EWcorot}$ fit from \cite{Nateghi2024_2} (C~\textsc{ii}, O~\textsc{vi}). Their $1\sigma$ uncertainty is $\sim0.1$ (see their Figure 4). In addition, we include the inner and outer CGM result from \cite{Kacprzak2025} (Mg~\textsc{ii}, O~\textsc{vi}). We scale their average inner (25 kpc) and outer (68 kpc) CGM bin choice by their average $\mathrm{R_{vir}}$ (129 kpc). 

At the center of our edge-on sample ($\mathrm{r/R_{200c}} \sim 0$), we find $f_\mathrm{EWcorot}\sim0.5$. In a symmetric disk viewed edge-on, we should expect the absorption feature to be symmetric about $\mathrm{0\,km\,s^{-1}}$. This is also seen in Figure \ref{fig:spectra}. For each ion, $f_\mathrm{EWcorot}$ reaches a maximum value at $\sim0.1\, \mathrm{r/R_{200c}}$ and slowly drops off as a function of distance along the midplane, in agreement with \cite{Nateghi2024_2} and \cite{Kacprzak2025}. The slope of the drop-off is independent of the ion. However, the value of $f_\mathrm{EWcorot}$ at a given impact parameter and z-height slightly decreases as a function of increasing ionization energy.   

\section{Discussion}\label{sec:discussion}

It is clear from \S\ref{sec:results} that the typical MW analog in IllustrisTNG has an extended, ionized, corotating structure. This is found in both the cool gas traced with H~\textsc{i} and Mg~\textsc{ii} and the warm gas traced with Si~\textsc{iv}, C~\textsc{iv}, and O~\textsc{vi}. We now compare our work to observations (\S\ref{subsec:comparison_to_observations}) and to other simulation work (\S\ref{subsec:comparison_to_simulations}).  We also discuss the role of satellites (\S\ref{subsec:satellites}) and the caveats of using IllustrisTNG for this study (\S\ref{subsec:caveats}).

\subsection{Comparison to Observational Work}\label{subsec:comparison_to_observations}

Much of the evidence for an extended, ionized disk is from external H~\textsc{i} and H$\alpha$ and its connection to the metagalactic background. We briefly discuss the H~\textsc{i} column density (\S\ref{subsubsec:hydrogen}). We then focus on co-rotation observed in Mg~\textsc{ii} (\S\ref{subsubsec:magnesium}), Si~\textsc{iv}, C~\textsc{iv}, and O~\textsc{vi} (\S\ref{subsubsec:oxygen}). Note that a large fraction of the observational literature is at $z\sim0.5$, whereas our analysis is at $z\sim0$, since we are interested in the present-day MW in a parallel HST/COS survey \citep{Werk2021hst,Werk2024}.    

\subsubsection{Neutral Gas (H~\textsc{i})}\label{subsubsec:hydrogen}

 
Our findings broadly agree with previous H~\textsc{i} observational work. For example, \cite{Kalberla2008} found that the MW H~\textsc{i} extends down to N(H)$\mathrm{\ \sim 10^{18}cm^{-2}}$ out to 60 kpc. The H~\textsc{i} disk of L$^\ast$ spirals has also been shown to extend out to $\sim$ $50 - 60$ kpc at N(H) $\sim$ $10^{18}$ cm$^{-2}$ \citep{Pisano2014,Wang2016, Wang2024, Wang2025}. We find that the median edge-on H~\textsc{i} column density of our sample is $\mathrm{\sim 10^{18}cm^{-2}}$ at $0.25\ \mathrm{R_{200c}}$ (54 kpc for our $\mathrm{R_{200c}}$ sample mean of 215 kpc). 
\cite{Bregman2018} suggest that the high H~\textsc{i} column density found in UV absorption could be explained by a large disk component at an impact parameter of 50 kpc. We find that a large fraction ($\gtrsim80\%$) of N(H~\textsc{i}) at $\sim50\,\mathrm{kpc}$ is aligned with the angular momentum axis of the galaxy (Figure \ref{fig:column_density}). At a larger impact parameter ($\sim 0.5\ \mathrm{R_{200c}}$) and lower column density (N(H~\textsc{i}) $\sim$ $10^{14}$ cm$^{-2}$), there is a non-negligible fraction of H~\textsc{i} along the galaxy major axis that is aligned with the angular momentum axis of the galaxy (Figure \ref{fig:column_density}). In fact, $\gtrsim 10\%$ of N(H~\textsc{i}) is aligned along the major axis out to the halo boundary, which is more than what we'd expect from a random distribution.

Our results also agree with recent work measuring the extent of the vertical diffuse H~\textsc{i}, which is likely driven by both feedback and gas accretion. In a sample mostly consistent with our stellar mass range, \cite{Yang2025} find that $z_{18}$, the extent of the N(H~\textsc{i}) $\sim$ $10^{18}$ cm$^{-2}$ contour along the minor axis, reaches out to $\sim$ $20-50$ kpc. We find that, for the typical MW analog, $z_{18}$ $\mathrm{\sim 0.1\,R_{200c}}$ ($\mathrm{\sim20}$ kpc) along the minor axis (see the minor axis profile in the upper panel of Figure \ref{fig:column_density}). We also note that our median subhalo H~\textsc{i} mass is $\mathrm{\log M_{H\text{\textsc{i}}}/M_\odot\sim10}$, which is consistent with their sample (see their Figure 7). \cite{Das2020,Das2024} find $z_{18}$ exceeding 50 kpc, but \citet{Yang2025} (see their Section 5.2) suggests caution when interpreting this detection

\cite{French_Wakker2020} calculate the $\mathrm{Ly\alpha}$ co-rotation fraction in a sample of sample of 33
galaxy–QSO pairs, where they define co-rotation using the difference between the absorber velocity and galaxy velocity \cite[similar to the binary approach of][]{Ho2017}. \cite{French_Wakker2020} find that $\sim85\%$ of the $\mathrm{Ly\alpha}$ absorption is corotating within 100 kpc and $\sim50\%$ outside 100 kpc (consistent with random motion).  
\cite{nateghi24} found that high H~\textsc{i} co-rotation ($f_\mathrm{EWcorot} \geq 0.50$) occurs along both the major and minor axes. They also found that co-rotation remains constant ($f_\mathrm{EWcorot}\sim0.6$) outside of the halo only along the major axis ($\phi \lesssim 30^\circ$). They suggested that this may be indicative of gas accretion along filaments along the major axis, from the intergalactic medium to the CGM.

We do not directly calculate $f_\mathrm{EWcorot}$ in H~\textsc{i}, but discuss $f_\mathrm{EWcorot}$ for other ions in the next section. \cite{nateghi24,Nateghi2024_2} suggested that H~\textsc{i} is likely tracing both the low ions (e.g., Mg~\textsc{ii}) and high ions (e.g., O~\textsc{vi}).   

\subsubsection{Ionized Cool Gas (Mg~\textsc{ii})}\label{subsubsec:magnesium}

The strong Mg II doublet (2796\AA) is typically observed at somewhat higher redshift (z $\sim$ 0.5) with ground-based studies, but it is useful to compare these results to our simulated MW galaxies. In general, there have been two approaches to studying Mg~\textsc{ii} around galaxies: (1) mapping out the geometry of Mg~\textsc{ii} absorption, and (2) probing the Mg~\textsc{ii} kinematics relative to the host galaxy. We focus on the latter. 

\cite{Steidel2002} originally found tentative evidence for Mg~\textsc{ii} rotation relative to the galaxy rotation curve at an impact parameter of $\sim 15-75$ kpc.  A similar conclusion was found in \cite{Kacprzak2010}, where 7 out of 10 systems had a line-of-sight velocity on one side of the galaxy systemic velocity with the same sign as the galaxy rotation curve.    
\cite{Ho2017} used a sample of 15 background quasars uniquely positioned near the midplane of foreground galaxies (z $\sim$ 0.2) and found similar extended Mg~\textsc{ii} co-rotation out to $0.35-0.45\ \mathrm{R_{vir}}$. 
\cite{Martin2019} followed up on this work and used a sample of 50 star-forming galaxies ($z \sim 0.2$), covering a larger range of azimuthal angles than \cite{Ho2017}. Within $46^\circ$ of the midplane, 12 systems were corotating, 5 had a rotation velocity consistent with the systemic velocity, and 3 did not have strong Mg~\textsc{ii} absorption.  \cite{Martin2019} suggest this hints at an asymmetric, \emph{less} flattened corotating CGM, especially with co-rotation at such a high azimuthal angle.  

In terms of angular momentum alignment, we find that that $\sim$ 50\% of the Mg~\textsc{ii} column density at $\sim 0.25\ \mathrm{R_{200c}}$ along the midplane is aligned with the angular momentum axis of the star-forming disk. At $\sim 0.5\ \mathrm{R_{200c}}$, the fraction of aligned material is only $\sim$ 10\% (Figure \ref{fig:corotation_fraction}). However, at $\sim 0.5\ \mathrm{R_{200c}}$, the median edge-on N(Mg~\textsc{ii}) $\sim10^{10}\ \mathrm{cm^{-2}}$, which is well below their detection limit. If we instead weight the co-rotation absorption fraction by the EW and filter out weak absorption, we find that the aligned absorption fraction is $\sim 0.6-0.8$ at $0.35-0.45\ \mathrm{R_{200c}}$ (Figure \ref{fig:absorption_fraction}). 

We now make a more direct comparison with observational work, where we calculate $f_\mathrm{EWcorot}$. We find agreement with \cite{Nateghi2024_2} and \cite{Kacprzak2025}, highlighted in Figure \ref{fig:Nateghi_comparison}.  In general, $f_\mathrm{EWcorot}$(Mg~\textsc{ii}) decreases outside of the inner $\sim0.2\,\mathrm{R_{200c}}$ and at larger z-height. In addition, $f_\mathrm{EWcorot}$(Mg~\textsc{ii}) also seems to approach the same value ($\sim0.6-0.7$) in the outer halo, independent of z-height. In fact, this is true for the other ions, too.

If we use the C~\textsc{ii} result \citep{Nateghi2024_2} as a proxy for Mg~\textsc{ii}, we find that our profile is $\sim0.1$ higher. However, \cite{Kacprzak2025} is nearly $\sim 0.20$ higher than our profile. This discrepancy is likely due to differences in (1) galaxy inclinations and (2) sightline positions. For example, our sightline sample is generated edge-on, whereas \cite{Nateghi2024_2} range from i = $0.1^\circ$ to i = $85.0^\circ$. \cite{Kacprzak2025} is limited to moderate inclination ($60^\circ\leq i \leq90^\circ$). 

In addition, our sightlines are confined to the midplane (out to $\mathrm{r/R_{200c}=0.75}$ and  $\mathrm{z/R_{200c}=0.5}$). In comparison, \cite{Nateghi2024_2} use quasars $>$ $1\, \mathrm{R_{vir}}$ and at a large azimuthal angle. \cite{Nateghi2024_2} only have 26 sightlines, 17 of which fall within our midplane region. In addition, \cite{Nateghi2024_2} define the quasar distance as the distance in projected two-dimensional space, \emph{not} distance along the projected midplane. Therefore, our result should be interpreted as an upper bound when compared to \cite{Nateghi2024_2}. \cite{Kacprzak2025} limit their azimuthal angle range to near the midplane ($1^\circ\leq \phi \leq35^\circ$), likely resulting in their higher co-rotation fraction.   

Lastly, we can ask, similar to \cite{Ho2017}: how many Mg~\textsc{ii} spectra are consistent with galaxy co-rotation? This comparison is not straightforward, since some work uses an absorption-selected sample and other work includes non-detections. In addition, some work defines systemic absorption, where the absorption is indistinguishable from the systemic velocity of the galaxy.

We calculate the following: (1) the fraction of sightlines where $f_\mathrm{EWcorot}>0.50$ with no detection limit, (2)  the fraction of sightlines where $f_\mathrm{EWcorot}>0.90$ with no detection limit, (3) the fraction of sightlines where $f_\mathrm{EWcorot}>0.50$ and $\mathrm{EW> 30\,m}$\AA, and (4) the fraction of sightlines where $f_\mathrm{EWcorot}>0.90$ and $\mathrm{EW> 30\,m}$\AA. The denominator of each fraction is the total number of sightlines, either including non-detections (1, 2) or not (3, 4). For (1), we find that $64\%$ Mg~\textsc{ii} sightlines in our sample (at any distance in our grid of $\sim50,000$ sightlines) are consistent with galaxy co-rotation. For (2), it is $50\%$. If we consider only observable Mg~\textsc{ii}, we find  $72\%$ and $55\%$ for (3) and (4), respectively. These results are consistent with observations. For example, \cite{Kacprzak2010} find that $70 \% $ (7 out of 10) systems are corotating with the galaxy. In addition, \cite{Ho2017}, find $67 \% $ (8 out of 12) and \cite{Martin2019} find $60\%$ (12 out of 20). 


We also make the same calculation for counter-rotation, where we now use either $f_\mathrm{EWcorot}<0.50$ or $f_\mathrm{EWcorot}<0.10$. We find that the counter-rotation fraction at any distance for (1) is $36\%$. For (2), it is $24\%$. If we account for the EW theshold, we find that the counter-rotation fraction is $28\%$ and $18\%$ for (3) and (4), respectively. There is no clear counter-rotation in \cite{Kacprzak2010}, where the remaining 3 galaxies have absorption velocities on both sides of the systemic velocity. \cite{Ho2017} do not find any counter-rotating systems. \cite{Martin2019} also do not find any counter-rotation near the midplane, but find nearly the same number of counter-rotating systems as corotating systems within $45^\circ$ of the minor axis. The discrepancy between our work and the literature may result from a combination of (1) our larger host galaxy sample, (2) our fixed inclination of $i=90^\circ$, and (3) our choice of using a sightline grid compared to the limited (and random) sightline positions in observational work.    

\subsubsection{Ionized Warm Gas (Si~\textsc{iv}, C~\textsc{iv}, O~\textsc{vi})}\label{subsubsec:oxygen}

From the MW perspective, \cite{Bish2021} found evidence of anomalously low levels of C~\textsc{iv} at high Galactic latitudes beyond $>10$ kpc from the MW's disk compared to the $\mathrm{L}^\ast$ population (also see \cite{Zheng2019}). They suggested that the bulk of the MW's CGM traced by C~\textsc{iv} is likely to be at low Galactic latitudes, or that the MW's CGM is lacking warm, ionized medium in general. When compared to our analyses on the MW analogs, the low-Galactic-latitude region in the MW is similar to what we define as the extended thick disk (Model 2, Region 1; Table \ref{table:corotation_fractions}). When we calculate the C~\textsc{iv} mass fraction, we indeed find that nearly 50\% of the C~\textsc{iv} CGM mass lies within this region and thus would not be detected with high-latitude sightlines, which is consistent with what is seen in the FOGGIE simulation \citep{Zheng2020}. 

It would be interesting to follow-up the archival N(C~\textsc{iv}) measurement presented in \cite{Bish2021} (see their Figure 6), but as a function of azimuthal angle. We find that there is a larger aligned fraction in the Model 3 extended disklike region, suggesting that aligned C~\textsc{iv} is found in a thick disklike structure. This would also support both the vertically and radially extended C~\textsc{iv} solutions proposed in \cite{Qu2022}.    

There has been a limited amount of Si~\textsc{iv} and C~\textsc{iv} in external galaxy work. In the COS-Dwarfs Survey, \cite{Bordoloi2014_CIV} found that C~\textsc{iv} is detected out to 100 kpc ($\sim0.5\ \mathrm{R_{vir}}$) and is bound to the dark matter halo. A similar result was found in \cite{Chen2001}, where there was a relatively steep drop-off in C~\textsc{iv} detection beyond 100 kpc. In comparison, \cite{Borthakur2013} find strong C~\textsc{iv} absorption out to 200 kpc (see Figure 10 in \cite{Bordoloi2014_CIV}). We find that the typical N(C~\textsc{iv}) profile in our sample is relatively shallow. $\log\,$N(C~\textsc{iv})$\ \sim13-14\ \mathrm{cm^{-2}}$ beyond $\mathrm{0.5\ R_{200c}}$, with no clear dependence on azimuthal angle. Observational work targeting C~\textsc{iv} has not looked at co-rotation.    

Instead, O~\textsc{vi} is more commonly used to measure extended co-rotation and its dependence on azimuthal angle, similar to Mg~\textsc{ii} (\S\ref{subsubsec:magnesium}). \cite{Kacprzak2015} found that the O~\textsc{vi} covering fraction was largest near the projected major axis ($\pm\ 10^\circ–20^\circ $) and projected minor axis ($\pm\ 30^\circ $). In follow-up work directly probing the line-of-sight velocity, \cite{Kacprzak2019} targeted a sample of 20 O~\textsc{vi}-absorption-selected galaxies ($0.15 < z < 0.55$) and found little evidence of co-rotation. There were 10 systems within $25^\circ$ of the major axis ($\sim0.15-0.65\ \mathrm{R_{vir}} $) and there were only 4 corotating systems. 1 of these 4 systems was also consistent with a disk rotation/accretion model. In addition, there were 5 systems consistent with counter-rotation. \cite{Kacprzak2019} suggested that this is a result of either 1. there is no strong kinematic signature in accreting gas (traced by O~\textsc{vi}) or 2. co-rotation and accretion is hidden by a large reservoir of diffuse O~\textsc{vi} with a velocity distribution within $\mathrm{\pm\ 200\ km\ s^{-1}}$.

More recently, \cite{Ho2025} came to a similar conclusion using new and archival data \cite[including][]{Kacprzak2019}. \cite{Ho2025} uniquely used systems where there were both low ions (Si~\textsc{ii}, Si~\textsc{iii}) and high ions (O~\textsc{vi}). Although there was no correlation found between O~\textsc{vi} and co-rotation, there was a correlation \emph{if} a low-ion match was present. \cite{Ho2025} suggest that this is indicative of co-spatial Si~\textsc{ii}, Si~\textsc{iii}, and O~\textsc{vi} in the inner CGM along the midplane. \cite{Ho2025} found that $\sim51 \% $ (21 out of 41) of the velocity components are corotating. We can compare to our work: how many O~\textsc{vi} spectra are consistent with galaxy co-rotation?  Using the same definitions outlined in \S\ref{subsubsec:magnesium}, we find that $67\%$ of O~\textsc{vi} is corotating ($f_\mathrm{EWcorot} > 0.50$) with the galaxy at any distance (1). For the more stringent co-rotation selection ($f_\mathrm{EWcorot} > 0.90$), we find that the fraction drops to $16\%$ (2). If we only include spectra where $\mathrm{EW> 30\,m}$\AA, we find that the co-rotation fraction is $70\%$ and $15\%$ for (3) and (4), respectively. If we consider $f_\mathrm{EWcorot} > 0.50$ and $\mathrm{EW> 30\,m}$\AA, we find that our results are higher than \cite{Kacprzak2019} and \cite{Ho2025}. This difference is similar to \cite{Kacprzak2019}, where they found more O~\textsc{vi} co-rotation in their simulations than their observations (see \S \ref{subsubsec: Warm Gas}).


For counter-rotating O~\textsc{vi}, we find that the fraction is $33\%$ and $4\%$ for (1) and (2), respectively. These are lowered to $30\%$ and $3\%$ for (3) and (4), respectively. \cite{Ho2025} emphasize that once the column density-weighted velocity is considered, there are nearly no counter-rotating systems. This is consistent with our work if we compare it to systems where a majority of the absorption signal ($f_\mathrm{EWcorot}< 0.10$) is counterrotating ((2) and (4)), as $>90\%$ of the absorption will have a Doppler sign opposite the galaxy's rotation. \cite{Kacprzak2019} find $50\%$ (5 out of 10) of sightlines near the midplane are consistent with counter-rotation. Similar to our discussion in \S\ref{subsubsec:magnesium}, these discrepancies may be due to the low number and location of sightlines in  observational work.

\cite{Nateghi2024_2} also calculate equivalent width co-rotation fraction ($f_\mathrm{EWcorot}$) in their O~\textsc{vi} absorption sample. Although \cite{Nateghi2024_2} did not separate O~\textsc{vi} with a low-ion match, they find that the median $f_\mathrm{EWcorot}$(O~\textsc{vi}) has a shallow slope, decreasing further from the galaxy. In addition, \cite{Nateghi2024_2} find that $f_\mathrm{EWcorot}$(O~\textsc{vi}) is more likely to reside along the major axis. A similar conclusion is found in \cite{Kacprzak2025}, where they compare sightlines in the inner CGM ($15-35$ kpc) and outer CGM ($35-100$ kpc). \cite{Kacprzak2025} find $f_\mathrm{EWcorot}$(O~\textsc{vi}) is $\sim$80\% ($\sim$60\%) in the inner (outer) CGM, which is consistent with our work. Similar to our discussion in \S\ref{subsubsec:magnesium}, discrepancies in our results and \cite{Nateghi2024_2,Kacprzak2025} likely arise from differences in inclination and sightline positions.  

\subsection{Comparison to Simulations}\label{subsec:comparison_to_simulations}

There has been a lot of work looking at the angular momentum evolution of the halo gas relative to the stellar disk and dark matter halo. We first briefly discuss the connection between angular momentum and our work (\S\ref{subsubsec: Angular Momentum}) and then extend our discussion to both the cool gas (\S\ref{subsubsec:Cool Gas})  and warm gas (\S\ref{subsubsec: Warm Gas}).

\subsubsection{Angular Momentum}\label{subsubsec: Angular Momentum}

There has been extensive simulation work on disk-formation \cite[see][for a review]{cgm-review-2023}. Broadly, it has been found that late-time gas accretion can result in filamentary gas with short cooling times that can form extended disklike structures \citep[e.g.,][]{Stewart2011,gutcke-gas-disks2017}. As a result, \cite{Stewart2011b} demonstrated that cold-mode accretion will result in halo gas orbiting with high angular momentum, resulting in an extended, warped, and corotating structure. \cite{Stewart2011b} named these structures `cold flow disks' and found that they tend to align with large-scale accretion along filaments. However, as discussed in \cite{Stewart2011b}, we may expect this co-rotation fraction to decrease towards lower redshift as cold-mode accretion \lq dies out'. In a simple calculation, we compute the net mass flow rate as a function of azimuthal angle (between the major and minor axes, see \S\ref{subsec:MW_misalignment}). If we consider the evolution of our MW sample since a redshift of z = 1, we find that (1) the major axis net mass flow rate is in the direction of the host galaxy (inflow) and dominant over the minor axis net mass flow rate (outflow) and (2) the net mass flow rate along the major axis has decreased since z = 1 (although still inflow). The structure of the inflowing gas is unclear (e.g.,  filamentary), but we find that it is dominated by $\mathrm{\sim 10^4\, K}$ gas. These findings are consistent with \cite{Stewart2011b}, but a more detailed analysis is outside of the scope of this paper and we defer it to future work. 

In a detailed study of MW-like halos in FOGGIE, \cite{Simons2024} looked at the angular momentum evolution of halo gas and dark matter relative to the stellar disk. They discussed the formation of an outer \lq secondary disk', refering to cold ($<1.5\times 10^4\ \mathrm{K}$) misaligned gas surrounding the central disk, created through filamentary accretion. Similar to our work, they defined misalignment as the angle between the cold gas net angular momentum in a given radial bin relative to the net angular momentum of the central stellar disk.  At low redshift ($\mathrm{z\sim 0}$), 3 out of 5 of their MW-like halos had a \lq secondary disk' misalignment $\lesssim 45^\circ$. The remaining 2 halos had a cold gas misalignment of nearly $180^\circ$ outside of the central $\sim 5\ \mathrm{kpc}$ (see their Figure 6).  It seems plausible that such outliers are similar to the $\sim 10\%$ of our sample which demonstrate a large fraction of counter-rotation in the halo.          

\cite{DeFelippis2020} also connected \lq cold' gas to angular momentum in a sample of IllustrisTNG MW-mass halos, where cold was defined as below half of the virial temperature, $\mathrm{T_{vir}}$ ($\mathrm{\sim 8\times 10^5\ K}$). This temperature range is not straightforward to compare to our work, since we expect Mg~\textsc{ii}, Si~\textsc{iv}, C~\textsc{iv}, and O~\textsc{vi} to \emph{all} be present in this temperature range. \cite{DeFelippis2020} found that the spatial distribution of high angular momentum \lq cold' gas is well-aligned with the galaxy stellar disk, largely corotating, and has an opening angle of $\sim30^\circ$. This wedge-like structure of the high angular momentum gas is similar to our aligned column density fraction (Figure \ref{fig:corotation_fraction}), especially in low ions. \cite{DeFelippis2020} also found this trend to a lesser extent in their \lq hot' gas ($\mathrm{\gtrsim 4\times 10^5\ K}$). The \lq hot' gas angular momentum offset is larger at a greater azimuthal angle compared to the  \lq cold' gas (see their Figure 4), similar to how we find an oblate-spheroid-like structure in warm gas (e.g., O~\textsc{vi}).      

\subsubsection{Cool Gas}\label{subsubsec:Cool Gas}

\cite{Stewart2011b} found that the \lq cold flow disks' should be offset from the galaxy systemic velocity by $\sim 100\ \mathrm{km\ s^{-1}}$ and have a velocity offset with the same sign as the galaxy rotation, as seen in the observational work discussed in \S\ref{subsec:comparison_to_observations}. In particular, \cite{Stewart2011b} look at inflowing gas where N(H~\textsc{i}) $\gtrsim10^{16}\ \mathrm{cm^{-2}}$, enabling a more direct comparison with Mg~\textsc{ii} absorption. From Figure \ref{fig:spectra}, we find that the typical Mg~\textsc{ii} absorption along the major axis has a velocity offset $\sim 50 - 150\ \mathrm{km\ s^{-1}}$ ($\bar{\mathrm{V}}_\mathrm{vir}=157\, \mathrm{km\ s^{-1}}$) in the direction of galaxy rotation. Beyond the center of the disk, there is negligible Mg~\textsc{ii} absorption counter-rotating or close to the systemic velocity of the galaxy.

There have been a few studies that examine the structure and kinematics of Mg~\textsc{ii} disk structure in simulated galaxies. \cite{Kacprzak2010} followed up on their sample of 20  Mg~\textsc{ii} absorption-selected galaxies (\S\ref{subsubsec:magnesium}) and examined Mg~\textsc{ii} absorption in a single galaxy ($\mathrm{z\sim 0.9}$). Similar to our work, they produced a grid of synthetic sight-lines and demonstrated that, independent of galaxy inclination ($i=0^\circ,\ 45^\circ,\ 90^\circ$), the Mg~\textsc{ii} velocity offset spans $\mathrm{\sim200\ km\ s^{-1}}$, with very little at the galaxy systemic velocity. In the average stack of Mg~\textsc{ii} absorption along the major axis, we find agreement with this statement, where the line-of-sight velocity is nearly always in the same direction as the galaxy rotation. One limitation in \cite{Kacprzak2010} was that they only used one galaxy. Although our typical galaxy may agree with their findings, there is galaxy-to-galaxy variation in our sample that is dependent on the cosmological environment that is not captured in their single galaxy.  

\cite{Ho2019} studied the extent of cold gas co-rotation in EAGLE, where the temperature range probed $\sim10^4\ \mathrm{K}$ is where we expect Mg~\textsc{ii} to exist. \cite{Ho2019} demonstrated that at low azimuthal angles ($\phi \lesssim 10^\circ$), a cold gas, corotating, disklike structure can be observed out to $\sim$ 60 kpc, supporting their earlier observational finding \citep{Ho2017}. This analysis was done with a single galaxy at $\mathrm{z=0.27}$ with a mass selection consistent with \cite{Ho2017} and \cite{Martin2019}. \cite{DeFelippis2021} also looked at Mg~\textsc{ii} absorption, but in a comparison between IllustrisTNG ($\mathrm{z \sim 1.0}$) and the MEGAFLOW Survey \citep{Zabl2019,Bouche2025}. In general, there was agreement between the simulation sample and MEGAFLOW Survey, supporting the model that there is co-rotation of the cold CGM near the galactic midplane. \cite{DeFelippis2021} also produced mock absorption spectra, similar to Figure \ref{fig:spectra}. At an impact parameter of 20 kpc, they found that the typical absorption is symmetric and centered at $\sim 0.60\ \mathrm{V_{vir}}$. At an impact of 40 kpc, the absorption is less symmetric, more noisy, and centered at $\sim 0.50\ \mathrm{V_{vir}}$. If we assume their MW analog sample has $\mathrm{R_{200c}\sim125\ kpc}$, then 20 kpc (40 kpc) is $\mathrm{\sim 0.15\ R_{200c}}$ ($\mathrm{\sim 0.30\ R_{200c}}$). At $\mathrm{\sim 0.15\ R_{200c}}$ (not shown in Figure \ref{fig:spectra}), we find Mg~\textsc{ii} centered at $\sim 0.9\ \mathrm{V_{vir}}$. At $\mathrm{\sim 0.30\ R_{200c}}$ (not shown in Figure \ref{fig:spectra}), we find Mg~\textsc{ii} centered at $\sim 0.6\ \mathrm{V_{vir}}$. Our sightlines in the inner disk have a higher line-of-sight velocity, likely owing to a difference in redshift and inclination. 

\subsubsection{Warm Gas}\label{subsubsec: Warm Gas}

\cite{Kacprzak2019} expanded upon their observational work (\S\ref{subsubsec:oxygen}) and analyzed O~\textsc{vi} in a suite of 8 simulated galaxies ($\mathrm{z\sim 1}$). In an averaged synthetic spectrum along the major axis, \cite{Kacprzak2019} found that there is O~\textsc{vi} absorption at $\sim100\ \mathrm{km\ s^{-1}}$ on either side of the galaxy, with the same sign as the galaxy rotation curve (see their Figure 9). In an O~\textsc{vi} velocity-weighted projection (see their Figure 8), they came to a similar conclusion: nearly all O~\textsc{vi} within $\sim 100\ \mathrm{kpc}$ should have rotation aligned with the galaxy rotation. This agrees with our work. However, it is difficult to make a direct comparison with their synthetic spectrum since their result is averaged over sightlines near the midplane ($\phi<30^\circ$), but extending out to $\mathrm{\sim200\, kpc}$.  

\cite{DeFelippis2020} found that the mass-weighted \lq cold' ($\lesssim4\times 10^5\ \mathrm{K}$) angular momentum wedge is associated with corotating gas. In particular, they found that \lq cold' gas along the midplane ($\mathrm{0 < z/R_{vir}<0.1}$) is $\sim 0.50\ \mathrm{V_{vir}}$ at $0.50\ \mathrm{R_{vir}}$ (see their Figure 11). We do not find a similar trend in our MW analog sample, where e.g., O~\textsc{vi} has typical rotational velocity $\sim 0.25\ \mathrm{V_{vir}}$ at $0.50\ \mathrm{R_{200c}}$ (Figure \ref{fig:spectra}). \cite{DeFelippis2020} also found similar alignment of the \lq hot' gas ($\gtrsim4\times 10^5\ \mathrm{K}$) angular momentum, where the inner half of the hot CGM is dominated by rotation. The hot gas along the midplane is rotating at $\sim 0.25\ \mathrm{V_{vir}}$ at $0.50\ \mathrm{R_{vir}}$. This is in better agreement with our O~\textsc{vi} result, likely owing to their choice of separating \lq cold' and \lq hot' where we expect O~\textsc{vi}. 

Our results, in combination with \cite{Kacprzak2019} and \cite{DeFelippis2020}, lend credence to the idea that the corotating kinematic structure of O~\textsc{vi} is not hidden (or absent) in simulation work. This is contrary to the observational work presented in \cite{Kacprzak2019} and \cite{Ho2025}, where they found no correlation with disk rotation. 
Instead, we find a clear co-rotation with the galaxy along the major axis, thus suggesting that Mg~\textsc{ii} and O~\textsc{vi} trace a similar underlying kinematic structure. It is unclear why there was very little O~\textsc{vi} co-rotation detected in \cite{Kacprzak2019}, but may have been a result of the small sample size. In future work, it would be interesting to see if there is a correlation between the low and high ions and their co-rotation in IllustrisTNG, similar to \cite{Ho2025}. 

We also note that the O~\textsc{vi} along the major axis that is corotating (and likely accreting) is largely photo-ionized, since it is tracing the cool filamentary inflow ($\sim 10^{4}\ \mathrm{K}$). However, O~\textsc{vi} can also have a  collisional ionization origin, and therefore be able to trace the hot halo \citep[$\sim 10^{5.5}\ \mathrm{K}$;][]{Oppenheimer2016}. O~\textsc{vii} is likely a more direct probe of the virialized hot halo ($\sim 10^{6-7}\ \mathrm{K}$), although the faint X-ray luminosity limits its observation to only our MW. \cite{HodgesKluck2016} measured O\textsc{vii} using both emission and absorption and found evidence of a rotating hot halo lagging behind the disk. They found that their data is inconsistent with a thick disk model compared to an extended halo model. However, it is unclear whether this structure is distinct from our work, since the corotating O~\textsc{vi} oblate-spheroid structure (Figure \ref{fig:corotation_fraction}) is similar to an extended halo. Since we don't make a distinction between O~\textsc{vi} photo- and collision ionization, is also possible that we are tracing the combination of a disklike component and extended halo, both of which are aligned with the angular momentum of the star-forming disk. We leave a more detailed analysis for future work. 

\subsection{Satellites}\label{subsec:satellites}

Satellites were intentionally excluded in our analysis. We now briefly investigate the role of satellites in shaping the angular momentum misalignment distribution in the CGM. We first calculate the azimuthal angle between every MW analog and satellite from the publicly available satellite catalog\footnote{\url{https://www.tng-project.org/data/milkyway+andromeda/}}. The azimuthal angle is the angle ($[0^\circ,90^\circ]$) between the host major axis and satellite. In our sample, there are 421 total satellites. 241 satellites are located within $30^\circ$ of the major axis. This is 57\%, which is larger than 50\% of the area of a sphere covered by $\mathrm{}b\pm30^\circ$ if the satellite distribution was uniform. This is broadly in agreement with the satellite overdensity along the major axis seen in both observational and simulation work \citep{Kang2007,Karp2023}. Naively, we may expect that the most massive satellites drive the angular momentum misalignment distribution in the CGM, since these systems bring in a large reservoir of gas. We therefore select satellites with a mass greater than the Large Magellanic Cloud ($\mathrm{M_{dyn}}>10^{11}\ \mathrm{M_\odot}$). We find that 7 out of the 11 LMC-mass systems are within $30^\circ$ of the host major axis. Each satellite belongs to a unique host\footnote{Subhalo IDs: 511303, 487742, 514272, 525533, 527309, 476266, 469487}.

We also calculate the net angular momentum axis of all gas in each satellite in the frame of the host MW analog (e.g., center of mass and bulk velocity of the host). In our sample, there are only 142 systems with gas present. Using our same definition of angular momentum alignment adopted earlier ($\theta\leq30^\circ$), we find that only 22 out of 142 satellites are aligned and 18 of these are within $30^\circ$ of the major axis. Only 1 out of the 7 most massive ($\mathrm{M_{dyn}}>10^{11}\ \mathrm{M_\odot}$) satellites within $30^\circ$ of the major axis are also aligned. 

We also look at the total satellite gas mass. In our sample, the median total satellite gas mass per MW analog is $\mathrm{\log M_g/M_\odot=9.15}$. 
This is calculated using all gas within the subhalo, if gas is present. If we look at the total gas mass distribution, we find that only 10\% of the total satellite gas mass in the MW analog sample is aligned with the host's angular momentum axis. This is similar to what we expect for a random distribution (\S\ref{subsec:define misalignment angle}). We can also look at the individual systems that compose the total satellite gas mass distribution. There are 90 systems with 0\% of their total gas mass aligned with the host galaxy angular momentum axis. In comparison, there are only 11 systems with $\geq 90$\% of their total gas mass aligned with the host galaxy angular momentum axis. 9 out of 11 are within $30^\circ$ of the major axis and 0 out of 11 are within the LMC-mass range. 

We also look at satellite-driven CGM counter-rotation. If we inspect the 10 MW analogs that are outliers (\S\ref{subsec:MW_misalignment}), we find 7 (2) MW analogs have their most massive satellite with $\theta \geq 90^\circ\, (150^\circ)$. Interestingly, nearly all counter-rotating CGMs only have satellites located at a low azimuthal angle. The 2 MW analogs with $\theta \geq 150^\circ$ each host a low-lying LMC-mass satellite.  

In conclusion, we find that there is an satellite overdensity near the host major axis, irrespective of satellite mass and distance. This is especially true for LMC-mass systems. In addition, if a satellite is aligned with the angular momentum axis of the host, it is more likely to be within $30^\circ$ of the host major axis, although this is not true for LMC-mass systems. However, there are not many systems that are aligned (only 22 out of 142). A similar conclusion is drawn when we look at the satellite gas mass distribution, not the satellite net angular momentum axis. We therefore shouldn't expect the gas from  satellites to contribute to our findings. In contrast, in counter-rotating MW analogs, the behavior is likely the result from a satellite interaction.          

It is also possible that both the co-rotation and counter-rotation is due to past satellite or filament accretion. We do not track the redshift evolution of satellites. It is possible that there is co-rotation or counter-rotation along the major axis that can be attributed to a satellite that has already merged or is currently not located near the major axis. This is beyond the scope of this work.

\subsection{Caveats of IllustrisTNG}
\label{subsec:caveats}

There are several important caveats to note when using IllustrisTNG. First, there is no radiative transfer, hence the local radiation field is not considered when using \textsc{trident} (\S\ref{subsec:trident}) to calculate the different ion species. However, as demonstrated using a simple \textsc{cloudy} model in Appendix \ref{appendix:stellar radiation field}, there is not a significant change in the expected ion properties outside of the immediate ISM. In fact, nearly all CGM ion modeling in the literature is done assuming only a uniform UV background. Second, IllustrisTNG resolution in the CGM of our MW analog sample is fairly coarse. The resolution is important, especially since the cold CGM may be fog-like and composed of sub-pc cloudlets \citep{McCourt2018,Gronke2020,Gronke2022}.

In addition, it has been shown that being able to resolve the interaction between the cold and hot phase of the CGM is critical for understanding the role of feedback \citep{Kim2020,schneider_physical_2020,Steinwandel2024}. This has motivated recent work to explore the use of a two-phase model in a cosmological context \citep{Fielding2022,Smith2024,smith_arkenstone_2024} and conduct a controlled experiment to increase the CGM resolution \citep{Hummels2019,Peeples2019,suresh_zooming_2019,Ramesh_Nelson2024}. For example, compared to our TNG50 sample with a CGM resolution of 1-3 kpc, \cite{Ramesh_Nelson2024} re-ran eight MW analogs in TNG50 with a CGM resolution of $\sim$ 300 pc. Despite this enhancement in resolution, the abundance of the smallest cold gas clouds has not converged. In addition, \cite{Ramesh_Nelson2024} found that the covering fraction of Mg~\textsc{ii} and C~\textsc{iv} do not converge numerically. This was not the case for O \textsc{vii}, which traces gas at $\sim 10^6$ K. Therefore, our predicted ion number density, especially in the low-to-intermediate ion range, is likely sensitive to our low CGM resolution. 

Interestingly, IllustrisTNG has a high baryon content in the CGM compared to other cosmological simulations (e.g., EAGLE), especially in the low-mass halo regime (see Figure 9 in \cite{Crain_VanDeVoort2023}). The IllustrisTNG feedback model does not efficiently heat and expand the surrounding medium, gas is not prevented from accreting into both the CGM and ISM, and a high mass loading factor is required to remove the gas from the ISM \citep{Wright2024}. Although we normalize by the total ion mass in Figure \ref{fig:ions}, other quantities we report (e.g., column density) may be under- or over-estimated.

It should also be noted that a stark difference was found between disk formation in MW analogs in IllustrisTNG and the original Illustris simulation. For example, it was found that neutral gas at large radii ($\sim100\ \mathrm{kpc}$) in IllustrisTNG is well-aligned with the galaxy major axis whereas it is more isotropic in Illusrtis \citep{Kauffmann2016,Kauffmann2019}. \cite{Kauffmann2019} found that this is a direct consequence of the different feedback implementation. The IllustrisTNG MW analog disks form from late-time cooling out of the hot halo whereas Illustris MW analog disks form from the accretion of cold, recycled material. In a sample of MW analogs in IllustrisTNG, \cite{Semenov2024b} found that there is both hot-mode inflow that can settle onto a disk and cold-mode accretion of clumpy material in the form of cold streams. We are likely capturing both of these mechanisms in our analysis. A more detailed redshift-dependent tracer particle analysis is needed. We leave this for future work.  

\section{Summary and Future Work}

\label{sec:conclusion}

We present evidence for an extended, ionized disk in a sample of 88 IllustrisTNG MW analogs. We quantify both the spatial and kinematic distribution of cool ($\sim 10^4\ \mathrm{K}$) and warm ($\sim 10^5\ \mathrm{K}$) halo gas, focusing on the ion species that trace this temperature regime: Mg~\textsc{ii}, Si~\textsc{iv}, C~\textsc{iv}, and O~\textsc{vi}. In particular, Mg~\textsc{ii} and O~\textsc{vi} are often observed in external galaxies and Si~\textsc{iv} and C~\textsc{iv} can be observed in our MW with HST/COS. We quantify both the fraction of gas aligned with the angular momentum axis of the disk \emph{and} the fraction of gas along a line-of-sight consistent with co-rotation (see \S\ref{subsec:agora} for clarification in our definition). This approach enabled us to bridge observationally motivated and theoretically motivated studies, linking the absorption of individual ions to their underlying angular momentum.

In our sample, the typical MW analog has an extended disklike structure in \emph{all} ions we model. We summarize our findings below: \\

\begin{itemize}
    \item In the $\sim10^4$ K gas (Mg~\textsc{ii}), angular momentum alignment is found in a disklike structure, but often doesn't extend beyond $0.20\ \mathrm{R_{200c}}$ (Figure \ref{fig:corotation_fraction}). The b/a ratio for the median edge-on Mg~\textsc{ii} angular momentum alignment fraction stack is 0.45 (0.49)  for 20\% (50\%) alignment (Table \ref{table:ba_ratio}). In the major (minor) axis region, 63\% (24\%) of the Mg~\textsc{ii} mass is corotating (Figure \ref{fig:ions}, Table \ref{table:corotation_fractions}).      
    
    \item In contrast, in the $\sim10^{5.5}$ K gas (O~\textsc{vi}), angular momentum alignment is found in an oblate-spheroid-like structure, extending well beyond $0.50\ \mathrm{R_{200c}}$ (Figure \ref{fig:corotation_fraction}). The b/a ratio for the median edge-on O~\textsc{vi} stack is 0.33 (0.56) for 20\% (50\%) alignment (Table \ref{table:ba_ratio}). 35\% (20\%) of the O~\textsc{vi} mass along the major (minor) axis is corotating (Figure \ref{fig:ions}, Table \ref{table:corotation_fractions})

    \item H~\textsc{i} is most similar to Mg~\textsc{ii}, where $\sim\ 64\%$ of the total H~\textsc{i} mass along the major axis is corotating. In comparison, Si~\textsc{iv} and C~\textsc{iv} have a spatial and kinematic distribution between Mg~\textsc{ii} and O~\textsc{vi}; a larger fraction of their total mass along the major axis is corotating compared to O~\textsc{vi} (Figure \ref{fig:ions}, Table \ref{table:corotation_fractions}).

    \item In a mock absorption line survey, we calculate the EW-weighted (and remove undetectable absorption) fraction of the total ion absorption from gas aligned with the angular momentum axis of the galaxy. In Mg~\textsc{ii}, we find $\gtrsim90\%$ of the absorption along the midplane within $\mathrm{r/R_{200c}\sim 0.2}$ is aligned with the angular momentum axis of the galaxy. The angular momentum alignment fraction decreases as an increasing function of z-height and ionization energy. $\sim60\%$ of the detectable O~\textsc{vi} at $\mathrm{r/R_{200c}\sim 0.2}$ along the midplane is aligned with the angular momentum axis of the galaxy  (Figure \ref{fig:absorption_fraction}).  
  
    \item In addition, we calculate the fraction of absorption consistent with galaxy rotation ($f_\mathrm{EWcorot}$), where we use the EW as a weight and select only absorption with $\mathrm{EW> 30\,m}$\AA. We find that $f_\mathrm{EWcorot}$ is consistent with recent observational work, where $f_\mathrm{EWcorot}$ decreases as a function of z-height and distance along the major axis. The relatively small difference we find may be due to the small observational sample and inclination (Figure \ref{fig:Nateghi_comparison}). In addition, we find that $\sim70\%$ of Mg~\textsc{ii} absorption is consistent with galaxy rotation ($f_\mathrm{EWcorot}>0.50$, $\mathrm{EW> 30\,m}$\AA). For O~\textsc{vi}, the sightline fraction is also $\sim70\%$ ($f_\mathrm{EWcorot}>0.50$, $\mathrm{EW> 30\,m}$\AA). Although Mg~\textsc{ii} is consistent with observational work that does not directly calculate $f_\mathrm{EWcorot}$, O~\textsc{vi} is higher (see \S\ref{subsubsec:magnesium} and \S\ref{subsubsec:oxygen}). We caution against a direct comparison since every work has a slightly different sample and co-rotation definition.    
    
    \item Lastly, we expect only a small fraction of the observed co-rotation to be from random motion or from satellite interactions. We find that only 10\% of the total satellite gas mass across the MW analog sample is aligned with the angular momentum axis of the central galaxy. In contrast, counter-rotating CGMs seem to be driven by massive counter-rotating satellites (see \S\ref{subsec:satellites}).
    
   \end{itemize}


In this study, we focus on the \emph{typical} MW analog. Future work is needed to explore this extended component on a galaxy-by-galaxy basis and to investigate how factors such as environment, merger history, and other physical processes influence the formation of the extended disklike structure in the CGM. Expanding this analysis to higher redshifts would offer further insight, especially when combined with the IllustrisTNG tracer particles, which trace the flow of gas.    

In upcoming work (Tchernyshyov, in prep.), we will compare the findings presented in this work to internal mock HST/COS sightlines, as well as to low latitude Si~\textsc{iv} and C~\textsc{iv} MW observations \citep{Werk2021hst, Werk2024}. Lastly, the Aspera SmallSat is set to launch in 2026 and will probe the diffuse O~\textsc{vi} emission around MW-like systems in the local Universe \citep{Chung2021}. It will be interesting to see if there is any clear signature of rotation in the CGM, especially since the instrument is not limited to the individual sight-lines used in absorption-line spectroscopy. This can also be combined with ongoing ground-based integral-field unit spectroscopy targeting a range of optical emission lines. The low surface brightness H\textsc{$\alpha$} emission is difficult to detect with current instrumentation, although there have been promising results from a recent stacking analysis \citep{Zhang2016,Zhang2018} and instrumentation development \citep{Melso2022,Melso2024}. 

\begin{acknowledgments}

M. Messere would like to thank Shy Genel and Mordecai-Mark Mac Low for the extensive feedback on the first-year report, which was the basis of this project.  We also thank Stephanie Ho for providing an early version of her most recent paper on O~\textsc{vi}.
GLB acknowledges support from the NSF (AST-2108470, AST-2307419), NASA TCAN award 80NSSC21K1053, and the Simons Foundation through the Learning the Universe Collaboration. This work was supported by NASA through grant HST-GO-16679 from STScI. JKW and KT also gratefully acknowledge support from NSF-CAREER 2044303.  
\end{acknowledgments}

\vspace{5mm}

\software{Astropy \citep{astropy}, \textsc{cloudy} \citep{cloudy}, \textsc{roco} \citep{ROCO}, \textsc{trident} \citep{trident}}

\appendix

\section{Stellar Radiation Field}
\label{appendix:stellar radiation field}

Here we describe the role of the Galactic radiation field on the spatial distribution of gas in our MW analog sample. Similar to other simulation analyses of the CGM, we assume the UV background that simulation was evolved with (see \S\ref{sec:methods}). However, this does not consider the potential effect of the local radiation field, which is not tracked in most cosmological simulations, including IllustrisTNG.  In order to investigate this, we model the radiation field as a spatial varying combination of a MW SED \citep{Fox2005} and UV background SED \citep{FG2009}, with self-shielding. We use the ionizing photon flux model in \textsc{galrad}\footnote{\url{https://github.com/deech08/galrad}} \citep{Barger2013,Hawthorn2019,Antwi-Danso2020} to calculate the \textsc{cloudy} intensity normalization for a given galaxy position. We select out to 100 kpc along the minor axis from the galaxy center in intervals of 5 kpc. This provides an upper limit to the photon flux, since we expect there to be a bipolar, ionization cone along the minor axis (see Figure 9 in \cite{Fox2005}). 

\begin{figure*}
\centering
\includegraphics[width=1\textwidth]{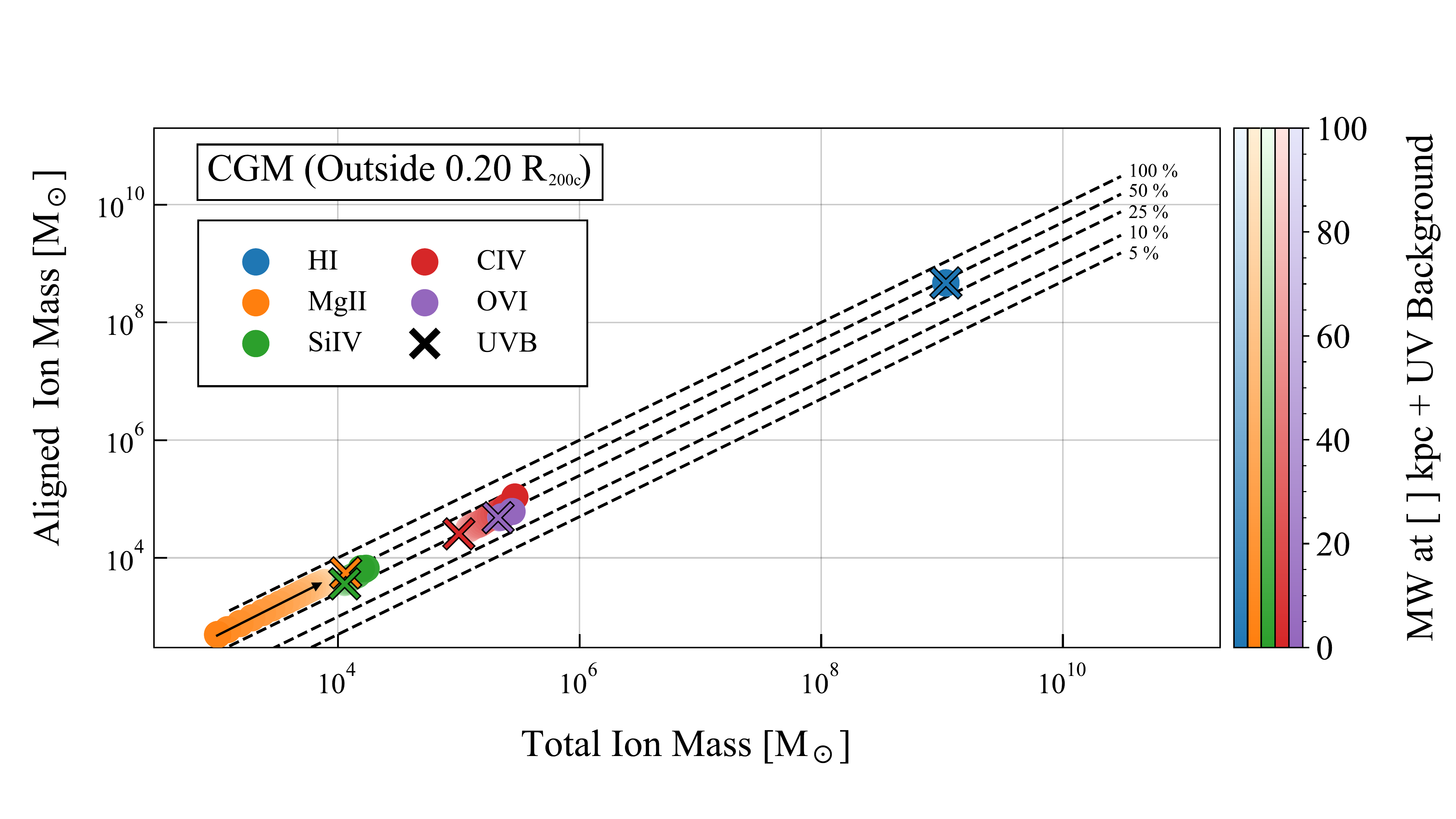}
\caption{The total and angular momentum aligned ion mass shift due to including the radiation field of the MW. At large radii, the radiation field will approach the UV background calculation used in this work. We include these using the \lq x' marker. As the radiation field of the MW begins to dominate at small radii (moving from a lighter shade to darker shade along the colorbar), the ions' mass will change. In particular, Mg~\textsc{ii} is the most sensitive to the inclusion of the MW radiation field. However, the ratio of the corotating mass to the total mass remains nearly the same.       
\label{fig:uv_background_evolution}}
\end{figure*}

We use \textsc{cloudy} \citep{cloudy} and \textsc{roco}\footnote{ \url{https://github.com/brittonsmith/cloudy_cooling_tools/tree/main}} \citep{ROCO} to calculate ion fractions for hydrogen number density and temperature that cover the range of values found in IllustrisTNG, assuming solar metallicity and abundances. In \textsc{cloudy} we use single-zone and iterate until convergence.  At 10 kpc from the Galactic center (not shown), the MW SED dominates over the UV background, especially in Mg~\textsc{ii}. However, beyond $\sim$35 kpc, the abundance fraction distribution slowly returns to the UV background only model. This is explicitly shown in Figure \ref{fig:uv_background_evolution}, where we calculate the ion total and corotating mass within the CGM as a function of  UV background and MW radiation field. In the CGM (outside $0.20\ \mathrm{R_{200c}}$), Si~\textsc{iv}, C~\textsc{iv}, and O~\textsc{vi} are not sensitive to the change in MW radiation field. H~\textsc{i} is negligible. However, the Mg~\textsc{ii} mass decreases by nearly a factor of 5 (denoted by the diagonal arrow in Figure \ref{fig:uv_background_evolution}). This conclusion is true for both the entire ion mass (horizontal axis) and angular momentum aligned ion mass (vertical axis). We also find that the evolution of the ratio between the aligned ion mass to the total ion mass is negligible. We included diagonal lines of constant mass ratios ($5\%,\,10\%,\,25\%,\,50\%,\,100\%,\,$) to highlight this. Note that we are \lq moving' all of the CGM mass along the minor axis, even though the radiation field should be different at each gas cell location in the halo. Therefore, our simple model can be viewed as an extreme case if the entire CGM was brought too close or too far away from the Galactic center. 

\cite{Shen2012} found that the local radiation field is negligible beyond $\sim$ 30 kpc. \cite{Zhu_Springel2024} developed a local photoionization model similar to \cite{Kannan2014}, where they take into account the radiation from both the star-forming regions and old stellar population and exclude the effect of the AGN. We also do not include the radiation field from the AGN feedback present in IllustrisTNG. They found a decrease in the H~\textsc{i} and Mg~\textsc{ii} column density and no change in in the O~\textsc{vi} column density beyond $0.3\ \mathrm{R_{200c}}$. In conclusion, they find that there is very little impact in the CGM, similar to our simple model presented here.              

\bibliography{sample631}
\bibliographystyle{aasjournal}

\end{document}